\documentclass[aps,prd,amsmath,amssymb,showpacs]{revtex4-2}

\usepackage{epsfig,float}
\usepackage{graphicx}
\usepackage{dcolumn}

\usepackage{morefloats}
\usepackage{color}
\usepackage{slashed}
\usepackage{bbm}
\usepackage{bm}
\usepackage{multirow}
\usepackage[colorlinks=true,linkcolor=blue]{hyperref}
\usepackage{subfigure}
\usepackage{appendix}

\begin{document}

\title{ On the widths of $\eta(1295)$ and $\eta(1405/1475)$ }

\author{ Yin Cheng$^{1,2}$\footnote{Email: chengyin@ihep.ac.cn}, Lin Qiu$^{1,2}$\footnote{Email: qiulin@ihep.ac.cn}, Qiang Zhao$^{1,2}$\footnote{E-mail:zhaoq@ihep.ac.cn} }

\affiliation{ 1) Institute of High Energy Physics,
        Chinese Academy of Sciences, Beijing 100049, P.R. China}

\affiliation{ 2) University of Chinese Academy of Sciences, Beijing 100049, P.R. China}

\begin{abstract}
Based on the assignment of the first radial excitation states of the isoscalar pseudoscalars for $\eta(1295)$ and $\eta(1405/1475)$, we investigate their three-body and four-body decay contributions to the total widths. In agreement with our previous studies we find that the triangle singularity (TS) mechanism arising from the intermediate $K^*\bar{K}$ rescatterings by exchanging a kaon or pion plays a crucial role in both $K\bar{K}\pi$ and $\eta\pi\pi$ channels. For the $\eta_X$ ($\eta_X$ stands for $\eta(1295)$ and $\eta(1405/1475)$) decays into $K\bar{K}\pi$, we find that although the transition $\eta_X\to K^*\bar{K}+c.c.\to K\bar{K}\pi$ is the dominant tree-level process, the productions of the intermediate $K\bar{\kappa}+c.c.$ and $a_0(980)\pi$ are strongly enhanced by the TS mechanism. For the $\eta_X$ decays into $\eta\pi\pi$, we find that the production of the intermediate $a_0(980)\pi$ via the triangle transition is the dominant one for $\eta(1295)$ partly because of the large $\eta(1295)K^*\bar{K}$ coupling. In contrast, the tree-level and triangle loop contributions are compatible and dominant in the $\eta(1405/1475)$ decays into $\eta\pi\pi$. It shows that a combined analysis is useful for disentangling the underlying dynamics for these two states. 

\end{abstract}

\maketitle 
 

\section{Introduction}

The nature of the states $\eta(1295)$, $\eta(1405)$ and $\eta(1475)$ has involved a lot of controversies during the past three decades. The main focus is whether there are two pseudoscalars ($\eta(1405)$ and $\eta(1475)$), or only one ($\eta(1405/1475)$) in the mass region of about $1.4\sim 1.5$ GeV.
Such a situation is due to several factors. Historically, the observation of a single $\eta(1405/1475)$ in $p\bar{p}$ annihilations at rest which decays into $K\bar{K}\pi$ has established this state as an excited isoscalar pseudoscalar~\cite{Baillon:1967zz}.
However, it was later found by MARK III~\cite{MARK-III:1990wgk} and DM-2~\cite{DM2:1990cwz} with increased statistics that the invariant mass spectrum seemed to favor a two-state structure around $1.44$ GeV. 
Such a solution was supported by the Obelix Collaboration at LEAR~\cite{OBELIX:2002eai}~\footnote{It should be pointed out that the fitted resonance parameters for $\eta(1405)$ and $\eta(1475)$ by MARKIII~\cite{MARK-III:1990wgk}, DM-2~\cite{DM2:1990cwz} and Obelix are not consistent~\cite{OBELIX:2002eai}.}.
As a consequence, the spectrum in the mass region of $1.3 \sim 1.5$ GeV which is anticipated to be the region for the first radial excitations states of $\eta$ and $\eta'$, appears with an overpopulation of states including $\eta(1295)$, $\eta(1405)$ and $\eta(1475)$.
Inspired by the phenomenological calculations of the ground state pseudoscalar glueball mass around $1.4$ GeV, and based on the interpretation that the $\eta(1405)$ seems to strongly couple to the $\eta\pi\pi$ channel while $\eta(1475)$ favors the $K\bar{K}\pi$ channel, the former is assigned as the candidate for the pseudoscalar glueball~\cite{Donoghue:1980hw,Close:1980rv,Close:1987er,Amsler:2004ps,Masoni:2006rz,Rosenzweig:1981cu,Cheng:2008ss,Close:1996yc,Li:2007ky,Gutsche:2009jh,Li:2009rk,Tsai:2011dp,Eshraim:2012jv} while the latter is assigned as the higher mass isoscalar pseudoscalar state of the first radial excitation nonet. 

However, the later lattice QCD simulations do not support a lower mass pseudoscalar glueball around 1.4 GeV.
Both quenched~\cite{Chen:2005mg,Bali:1993fb,Morningstar:1999rf,Chowdhury:2014mra} and unquenched calculations suggest that the pseudoscalar glueball mass should be around $2.4-2.6$ GeV. 
Meanwhile, more and more high-precision data from $J/ \psi$ and $\psi(3686)$ decays at BESII and BESIII reveal that it is sufficient to describe the invariant mass spectrum by one state though the mass positions could be slightly shifted in different channels.
 For instance,  the mass extracted in the $K\bar{K}\pi$ channel is $1452.7 \pm 3.8$ MeV~\cite{BESIII:2013cbb}, while those in $\eta \pi \pi$~\cite{BESIII:2011nqb,BESIII:2010gmv,BESIII:2019yzg} and $3\pi$~\cite{BESIII:2012aa} are about 1405 MeV. 
 Recent phenomenological studies based on the U(1)$_A$ anomaly dynamics also find that the lowest pseudoscalar glueball mass should be much higher than 1.4 GeV~\cite{Mathieu:2009sg,Qin:2017qes,Li:2021gsx}.

A breakthrough for this puzzling situation was triggered by the observation of the abnormally large isospin-breaking effects in $J/ \psi \to \gamma \eta(1405/1475) \to \gamma +3 \pi$~\cite{BESIII:2012aa}, which can be explained by the 
presence of the ``triangle singularity" (TS) mechanism~\cite{Landau:1959fi,Cutkosky:1960sp} due to the intermediate $K^*\bar{K}+c.c.$ rescatterings by exchanging an on-shell kaon (anti-kaon) in the decay of $\eta(1405/1475) \to 3 \pi$~\cite{Wu:2011yx}. 
Assuming that the $\eta(1405)$ and $\eta(1475)$ signals are originated from an identical state around $1.40\sim 1.44$ GeV~\footnote{In this work we refer the single state as either $\eta(1405/1475)$ or just $\eta(1405)$ if it does not bring confusions.}, the TS mechanism can naturally explain the mass shift and decay patterns observed in experiment~\cite{Wu:2011yx,Wu:2012pg,Aceti:2012dj}.
 A followed-up series of studies have explored broadly the role played by the TS mechanism in the productions and decays of these two states which advanced our understanding of the nature of $\eta(1295)$ and $\eta(1405/1475)$~\cite{Du:2019idk,Cheng:2021nal,Achasov:2015uua,Liu:2015taa,Achasov:2021yis,Nakamura:2022rdd}.
  Meanwhile, further studies are still needed to test such a scenario and help us gain more insights into the pseudoscalar spectrum.

To further demonstrate the nature of the $\eta(1295)$ and $\eta(1405/1475)$ as conventional quark model states, which however are strongly affected by the TS mechanism, in this paper, we systematically study the total widths and the related partial widths of these two states based on the first radial excitation assumption.
We have included the contributions of all the possible intermediate channels such as $\bar{K}K^*+c.c.$, $\kappa \bar{K}+c.c.$, $a_0(980) \pi $, $\sigma \eta $, and $f_0(980) \eta$. All these channels will feed in either $K\bar{K}\pi$ or $\eta\pi\pi$ which are the final states measured in experiment. To some extent, the decay modes of the isoscalar  pseudoscalars are rather simple. Although the intermediate channels are various, we will see that the combined analysis is informative to extract some featured dynamics for understanding their properties. One also notices that the present experimental data on $\eta(1405)$ and $\eta(1475)$~\cite{Workman:2022ynf} are rather confusing and the status is far from satisfactory. It shows that our analysis is able to provide a coherent picture for most of these existing data based on the first radial excitation scenario for $\eta(1295)$ and $\eta(1405/1475)$ associated by the TS mechanism.

To proceed, we introduce the formalism for $\eta_X$ ($\eta_X$ denotes $\eta(1295)$ or $\eta(1405)$) decays in an effective Lagrangian approach in Sec.~\ref{sec:2}. In Sec.~\ref{sec:3}, we present the numerical results in comparison with the experimental data and discuss their phenomenological consequences. A brief summary is given in Sec.~\ref{sec:4}.

\section{Formalism}\label{sec:2}

\subsection{Experimental status}
   
As mentioned in the Introduction that the experimental status of $\eta(1405)$ and $\eta(1475)$ is rather confusing and sometimes appears to be inconsistent, it is necessary to provide a brief summary of the experimental data for $\eta(1295)$, $\eta(1405)$ and $\eta(1475)$. In  Table~\ref{Tab:data1405} and~\ref{Tab:data1295} we list the averaged data for $\eta(1295)$ and $\eta(1405/1475)$ from Particle Data Group (PDG)~\cite{Workman:2022ynf}, respectively. 
In PDG, the mass and width of $\eta(1295)$ are $1294\pm4$ MeV and $55\pm 5$ MeV, respectively.
While some experiments obtained a larger width, such as the E852 collaboration measurement is $66\pm 13$ MeV~\cite{E852:2000rhq} and the measurement in Ref.~\cite{Stanton:1979ya} is about $70$ MeV.  
Note that the data of $\eta(1295)$ are lesser than that of $\eta(1405)$ due to the suppression on the production of $\eta(1295)$ in the $e^+ e^-$ collisions via $J/\psi\to\gamma \eta(1295)$~\citep{Wu:2011yx,Wu:2012pg}.
  
The mostly studied channels for the $\eta(1405/1475)$ decays are the three-pseudoscalar final states, namely, $K\bar{K}\pi$ and $\eta \pi \pi$. The relevant intermediate two-body decay channels can be $K^*\bar{K}+c.c.$, $\kappa \bar{K}+c.c.$, $a_0(980)\pi$, $\sigma/f_0(980)\eta$, etc. However, the present experimental information about these intermediate states are far from satisfactory. In particular, the contributions from the broad scalars are rather unclear. 
In Ref.~\cite{Du:2019idk}, a combined analysis of $\eta(1405)$ decays to $K\bar{K}\pi$, $\eta \pi \pi$ and the isospin violating channel $3 \pi$ further clarifies the important role played by the TS mechanism. It is also shown in Ref.~\cite{Du:2019idk} that these two channels cannot saturate the total width of $\eta(1405)$. In experiment, the partial wave analysis on the $K\bar{K}\pi$ spectrum shows that the $S$-wave $K \pi$ (i.e. $\kappa$) contributions cannot be ignored~\cite{BES:1998bgh,BES:2000adm,MARK-III:1990wgk}. 

It is necessary to discuss the reliability of the data.  As shown in Table.~\ref{Tab:data1405}, the branching ratio (B.R.) fraction of $ \Gamma(\eta(1405)\to \eta \pi \pi)$ to $ \Gamma(\eta(1405)\to K \bar{K} \pi)$ measured in $p \bar{p}$ collision is $1.09 \pm 0.48$~\cite{Amsler:2004rd}, while the ratio of $BR(J/\psi \to \gamma \eta(1405) \to \gamma \eta \pi^+\pi^- )$ over $BR(J/ \psi \to \gamma \eta(1405) \to \gamma K \bar{K} \pi )$ measured by BESII is just $0.16 \pm 0.04$~\cite{BES:1998bgh,BES:2000adm,BES:1999axp,Workman:2022ynf}. 
In Ref.~\cite{BES:1998bgh}, it is pointed out that there are two reasons accounting for such a large deviation. 
First, in $p\bar{p}$ collision, the analysis did not incorporate the interference between $\eta(1405)$ and a higher pseudoscalar $\eta(1800)$. Secondly, due to the phase space available for the $\sigma$ is limited in $p\bar{p} \to \eta(1405) \sigma$, the partial width of $\eta(1405)\to K^* \bar{K}$ will be weaken. While such suppressions are absent in the $J/\psi$ decays, we choose the $J/\psi$ decay data as input to constrain the relevant parameters in our analysis.

In the PDG~\cite{Workman:2022ynf} $\eta(1405)$ and $\eta(1475)$ are listed as two different states based on the two resonance scenario. For instance, the PDG categorizes that the $\eta(1405)$ dominantly decays into $\eta\pi\pi$ in various processes. But data from BESIII for its decays into $K\bar{K}\pi$ are not included. The obvious consideration is that the $0^{-+}$ resonance around 1.4 GeV in the $K\bar{K}\pi$ channel generally has a slightly higher mass. Thus, it is categorized as $\eta(1475)$. Surprisingly, the BESIII data for $\eta(1405/1475)$ in $J/\psi\to\gamma \eta\pi\pi$ and $\gamma K\bar{K}\pi$ have not been included in the PDG evaluations. The inconsistency between the PDG categorizing and the high-statistics data is that except for the $K\bar{K}\pi$ channel all the other decay channels for the $0^{-+}$ resonance structures around 1.4 GeV in charmonium decays do not indicate two resonances simultaneously in any exclusive decay channels, such as $J/\psi \to \gamma\eta\pi\pi$, $\gamma 3\pi$, $\omega\eta\pi\pi$, and $\gamma\gamma\phi$, etc.  Some of those two-state analysis may even cause misunderstandings. As shown in  Table.~\ref{Tab:data1405} the ratio $\Gamma(K^*\bar{K}\to K\bar{K} \pi) /\Gamma(a_0(980)\pi \to K\bar{K} \pi)$ for $\eta(1405)$ turns out to be very small~\cite{E852:2001ote}. However, it is because most contributions to the $K^*\bar{K}$ have been assigned to the so-called $\eta(1475)$.

In this analysis we aim at a survey of the main mechanisms which contribute to the total widths of $\eta(1295)$ and $\eta(1405)$. Apart from the intermediate $K^*\bar{K}+c.c.$ and $a_0(980)\pi$ channels, we also include other scalar-pseudoscalar intermediate states which contribute to the three-pseudoscalar final states, such as $\sigma/f_0(980)\eta$ and $\kappa \bar{K}+c.c.$ In addition, we also investigate the four-pseudoscalar final state decays such as $K\bar{K}\pi\pi$ and $4\pi$ which can contribute via the productions of the intermediate $K^*\bar{K}^*$ and $\rho\rho$. We note that the three pion decay channel is isospin-violating and does not contribute to the total width significantly. But the analysis on the $3 \pi$ channel similar to that in Ref.~\cite{Du:2019idk} will offer a reasonable constrain on the parameters.

\begin{table}
  \caption{Available data for $\eta(1405/1475)$ from the PDG~\cite{Workman:2022ynf} and theoretical results from two fitting schemes. In the column labelled by ``Notes'' we note the main features of the data related to either ``two resonances (T.R.)" or ``one resonance (O.R.)" solutions which are used in the partial wave analysis in the mass region around $1.4\sim 1.5$ GeV. Experimental data adopted as input for the determination of model parameters in Scheme-I are highlighted in boldface. }\label{Tab:data1405}
  \centering
  \begin{tabular}{lllcc}
    \hline \hline
      Observables    &   Experimental data                                                                                                      & Notes                        & Scheme-I                  & Scheme-II \\
    \hline
   \multirow{2}{*}{ $\Gamma(\eta \pi \pi )/\Gamma(K\bar{K}\pi)$}                            &     ${\bf 0.16 \pm 0.04}$ \cite{Workman:2022ynf}            &  $J/\psi$ decays         &   \multirow{2}{*}{$0.18\pm0.02$}     &  \multirow{2}{*}{$0.20\pm0.24$}   \\
                                                                                            &     $1.09 \pm 0.48$ \cite{Amsler:2004rd,Baillon:1967zz}     &  $p\bar{p}$ collision         &                                        &  \\    \hline                                                                            
     
   $\Gamma(K^*\bar{K}\to K\bar{K} \pi) /\Gamma(a_0(980)\pi \to K\bar{K} \pi)$               &    $0.08\pm 0.02$ \cite{E852:2001ote}                     &  T.R.: $\eta(1405)$           &    $25.20\pm2.80$                       &    $25.20\pm13.20$  \\
 
   $\Gamma(a_0(980)\pi \to K \bar{K} \pi )/ \Gamma(K\bar{K} \pi )$                           &     $\sim 0.15$ \cite{OBELIX:1995zjg}                       &  T.R.: $\eta(1405)$           &   $(2.80\pm 0.30)\times 10^{-2}$       &   $0.02\pm0.01$   \\  \hline

   \multirow{2}{*}{$\Gamma(K^*\bar{K}\to K\bar{K} \pi ) / \Gamma(K \bar{K} \pi )$}                                 &    $0.50 \pm 0.10$ \cite{Baillon:1967zz}                                  &   O.R.                     & \multirow{2}{*}{$0.67\pm0.05$}  &   \multirow{2}{*}{$0.49\pm0.16$}     \\
                                                                                                                   &    ${\bf 0.70 \pm 0.05}$  \cite{BES:2000adm}                              & O.R.                       &                                   &\\  \hline 

  \multirow{2}{*}{$\Gamma(K^* \bar{K}\to K\bar{K}\pi)$ /$\Gamma(\kappa \bar{K} \to K \bar{K} \pi)$  }             &     $2.70$ \cite{BES:1998bgh}                              & O.R.                           & \multirow{2}{*}{ $1.20\pm0.09$ }   &  \multirow{2}{*}{ $0.78\pm0.30$ } \\
                                                                                                                  &     $5.40 \pm 1.30$ \cite{BES:2000adm}                      & O.R.                           &                                    &  \\
    \hline                                                                                                   
 
   \multirow{3}{*}{$\Gamma(a_0(980)\pi \to \eta \pi \pi ) / \Gamma(\eta \pi \pi )$ }                          &       $0.29 \pm 0.10$ \cite{CrystalBarrel:1998pap}           & O.R. &  \multirow{3}{*}{$0.60\pm0.08$}     &  \multirow{3}{*}{$0.55\pm 1.00$} \\
                                                                                                              &       $0.19 \pm 0.04$ \cite{GAMS:1997pxg}                    & --&   & \\ 
                                                                                                              &       ${\bf 0.56\pm 0.04\pm 0.03}$ \cite{CrystalBarrel:1995kfe}    & O.R.& &\\
    
    \hline          
      $\Gamma(\eta (\pi \pi )_{S-\text{wave}})/\Gamma(\eta \pi \pi )$                           &    $0.81\pm 0.04$ \cite{GAMS:1997pxg}                                      &-- &   $0.15\pm0.02$  &  $0.21\pm0.30$ \\
      $\Gamma(f_0(980)\eta \to \eta \pi \pi )/ \Gamma(\eta \pi \pi )$                           &    $0.32 \pm 0.07$ \cite{Anisovich:2000kx}                                 & O.R. & $0.14\pm0.02$  &  $0.11\pm0.14$\\
    \hline
    \multirow{3}{*}{ $\Gamma(a_0(980)\pi \to \eta \pi \pi) / \Gamma(\eta (\pi \pi )_{S-\text{wave}})$}      &    $0.91\pm 0.12$ \cite{Anisovich:2001jb}                      & O.R. &      \multirow{3}{*}{$3.93\pm0.48$}  &  \multirow{3}{*}{$2.62\pm4.44$} \\
                                                                                                            &    $0.15\pm 0.04$ \cite{E852:2000rhq}                          & O.R. &   &\\
                                                                                                            &    $0.70\pm 0.12 \pm 0.20$ \cite{BES:1999axp}                  & O.R.&    &\\
    \hline         
     $\Gamma_{\eta(1405)}$                                                                                  &       ${\bf 50.10 \pm 2.60}$  \cite{Workman:2022ynf}                   &     &   $52.49\pm4.41$   & $42.30\pm13.10$ \\    
    
    \hline \hline
  \end{tabular}
 
\end{table}

\begin{table}
  \caption{Available data for $\eta(1295)$ from the PDG~\cite{Workman:2022ynf} and the theoretical predictions from two fitting schemes.}\label{Tab:data1295}
  \centering
  \begin{tabular}{lccc}
    \hline \hline
    Branching ratios      &    PDG averaged data                                                                                       & Scheme-I  & Scheme-II    \\
    \hline 
     $\Gamma( a_0(980)\pi\to \eta \pi \pi )/ \Gamma( \eta \pi^0 \pi^0 )$      \qquad & \qquad  $0.65 \pm 0.10$ \cite{GAMS:1997pxg}     &  $0.89\pm0.08$                          &  $0.78\pm0.73$     \\
     $\Gamma( a_0(980)\pi) / \Gamma( \sigma \eta)$                           \qquad  & \qquad  $0.48\pm 0.22$ \cite{E852:2000rhq}      &  $229\pm 42$                            &  $48.40\pm67.30$      \\
     $\Gamma( \eta (\pi \pi )_{S-\text{wave}})/ \Gamma( \eta \pi^0 \pi^0)$    \qquad & \qquad  $0.35 \pm 0.10$  \cite{GAMS:1997pxg}    &  $(4\pm 1)\times 10^{-3}$          &  $(1.60\pm2.20)\times 10^{-2}$   \\
     $\Gamma_{\eta(1295)}$                                                           &         $ 55.00 \pm 5.00$ MeV \cite{Workman:2022ynf}  &  $91.69\pm 6.12$ MeV                     &  $69.80\pm32.30$     \\
     \hline \hline
  \end{tabular}
 \end{table}

\subsection{Three-pseudoscalar decay mechanisms}

\begin{figure}
  \centering
  \subfigure[]{ \includegraphics[width=1.6in]{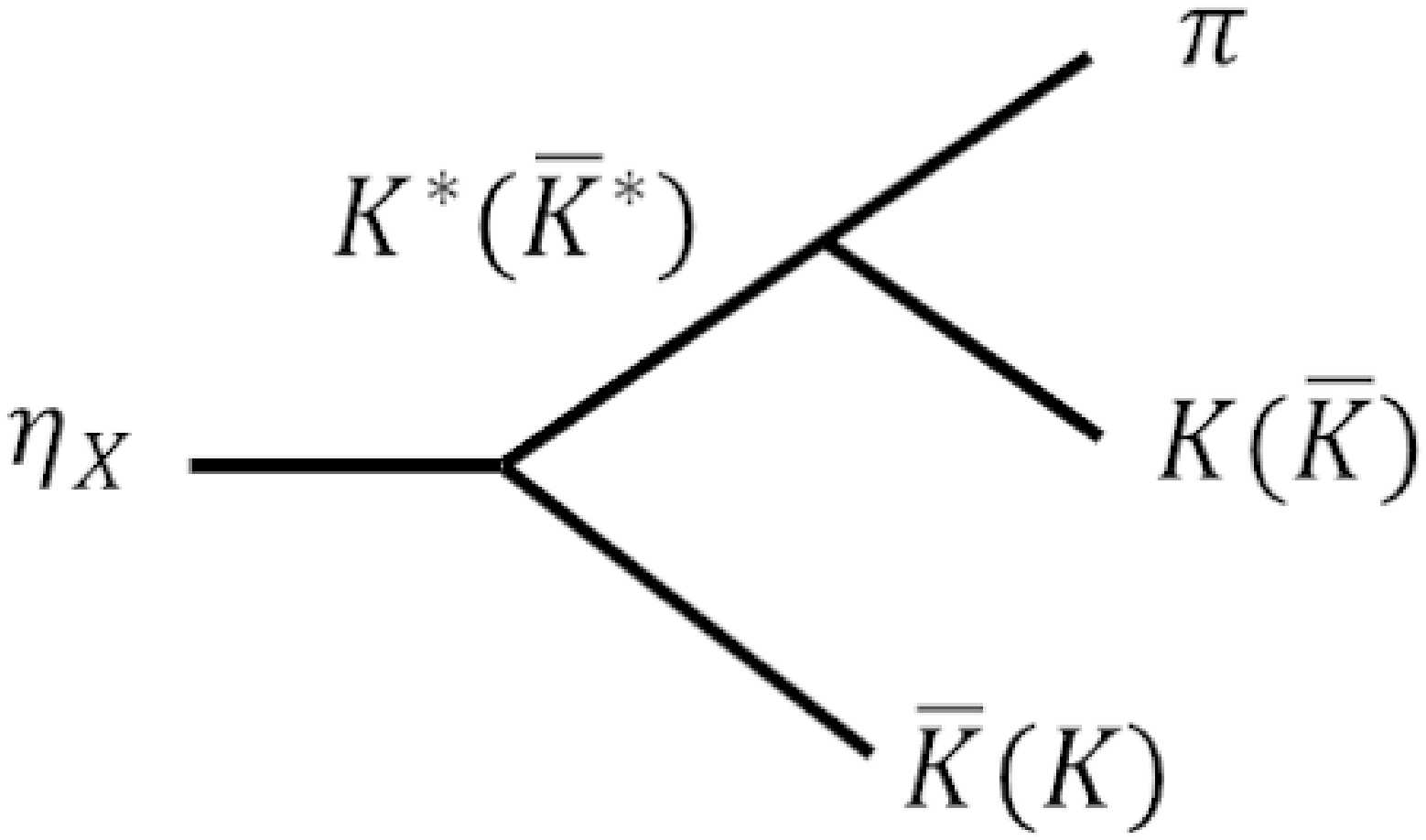}} \qquad
  \subfigure[]{ \includegraphics[width=1.5in]{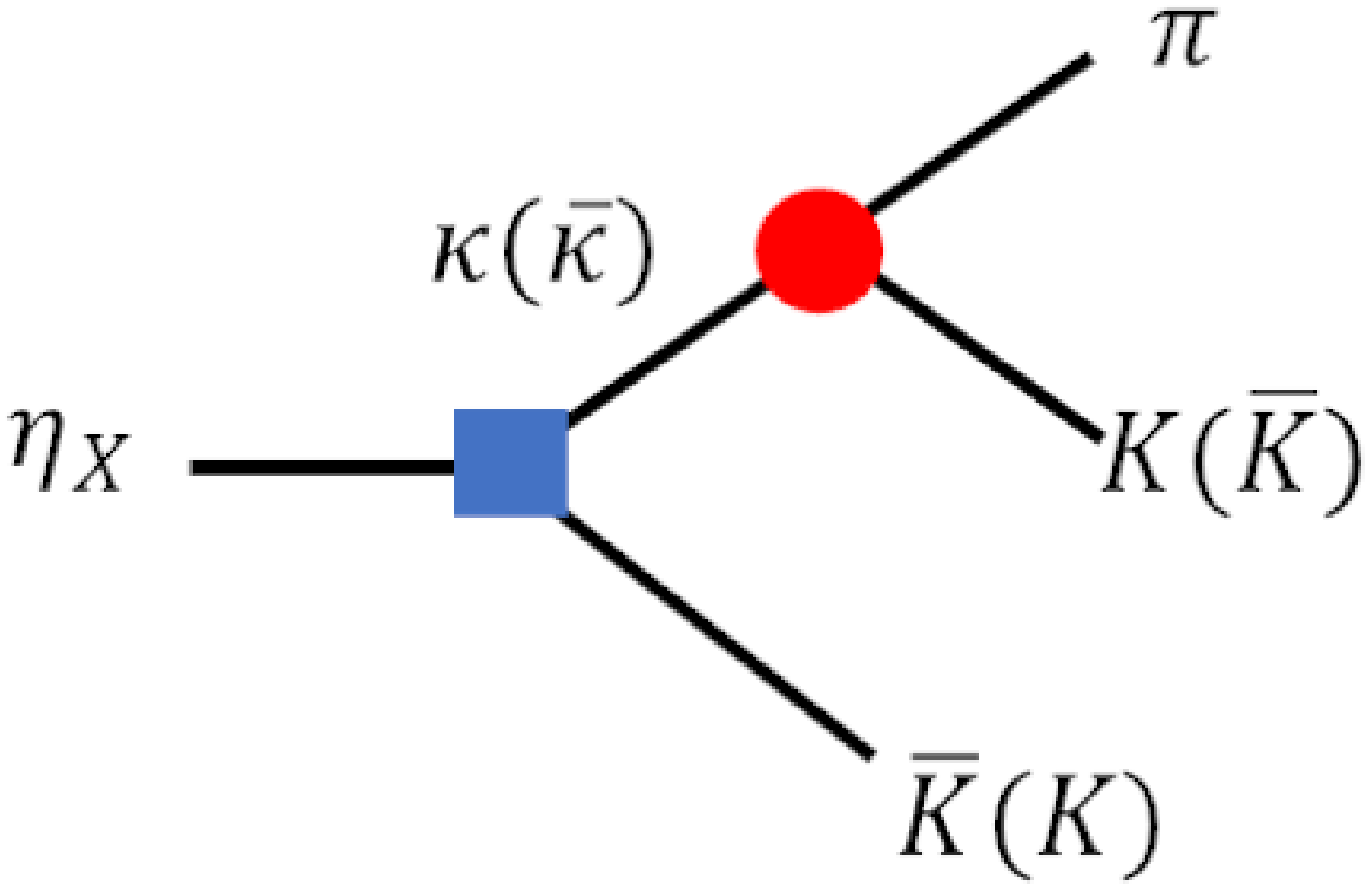}} \qquad
  \subfigure[]{ \includegraphics[width=1.4in]{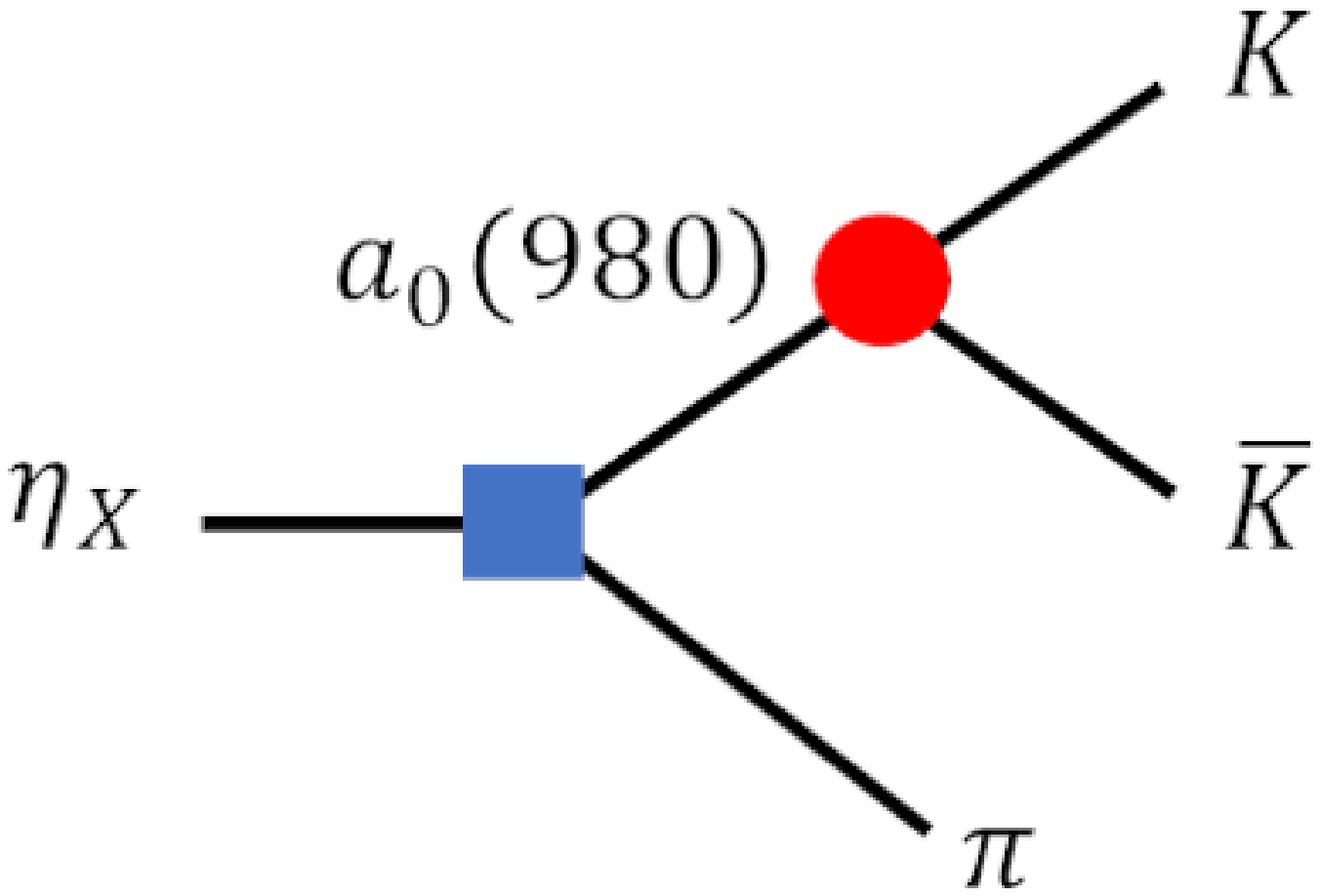}}  \\
  \subfigure[]{ \includegraphics[width=1.4in]{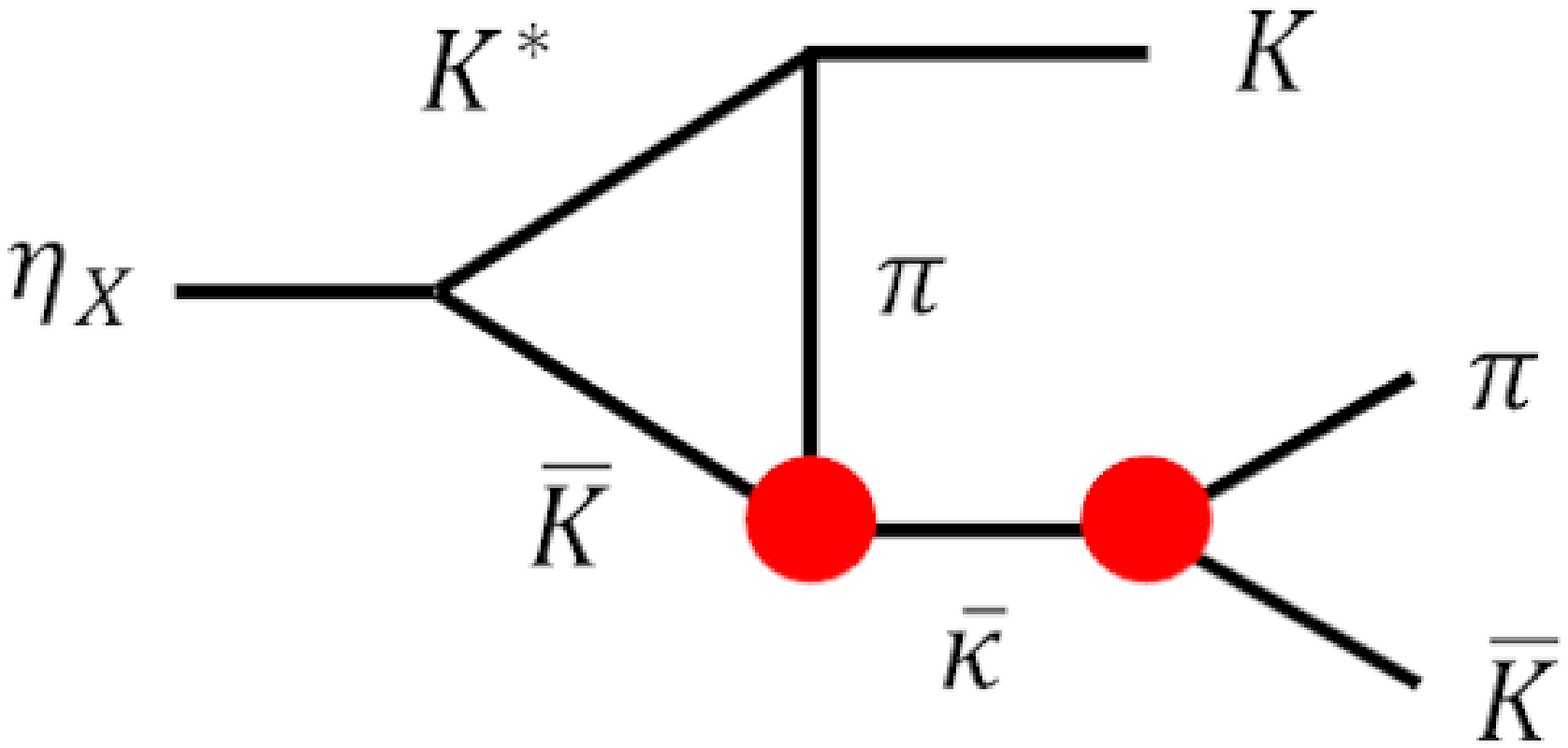}}  \qquad 
  \subfigure[]{ \includegraphics[width=1.4in]{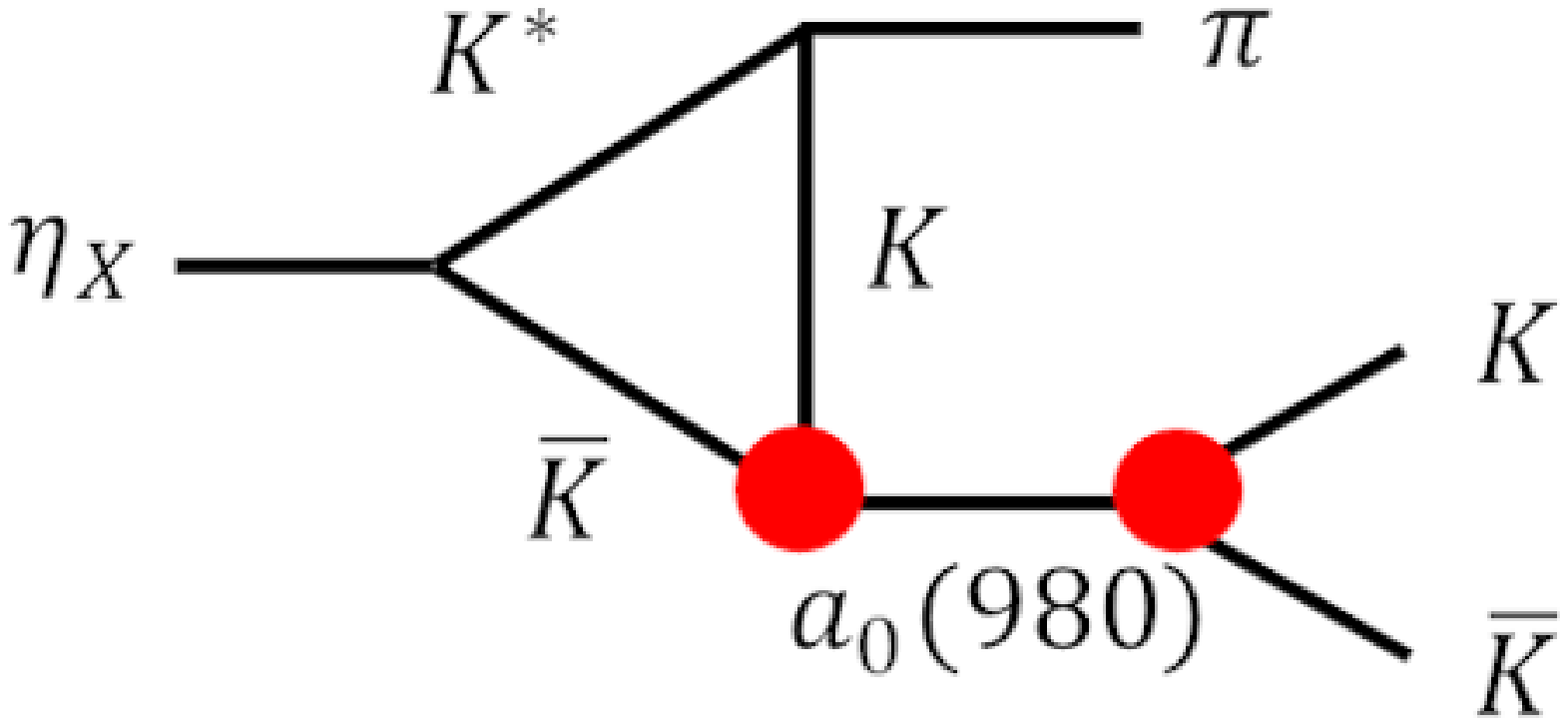}}
  \caption{Tree level and one-loop diagrams for $\eta_X \to K \bar{K} \pi.$ The blue solid squares indicate the $XSP$ couplings and the red solid circles indicate the $SPP$ couplings.}\label{fig:kkpidiagram}
\end{figure}

\begin{figure}
  \centering
  \subfigure[]{\includegraphics[width=1.5in]{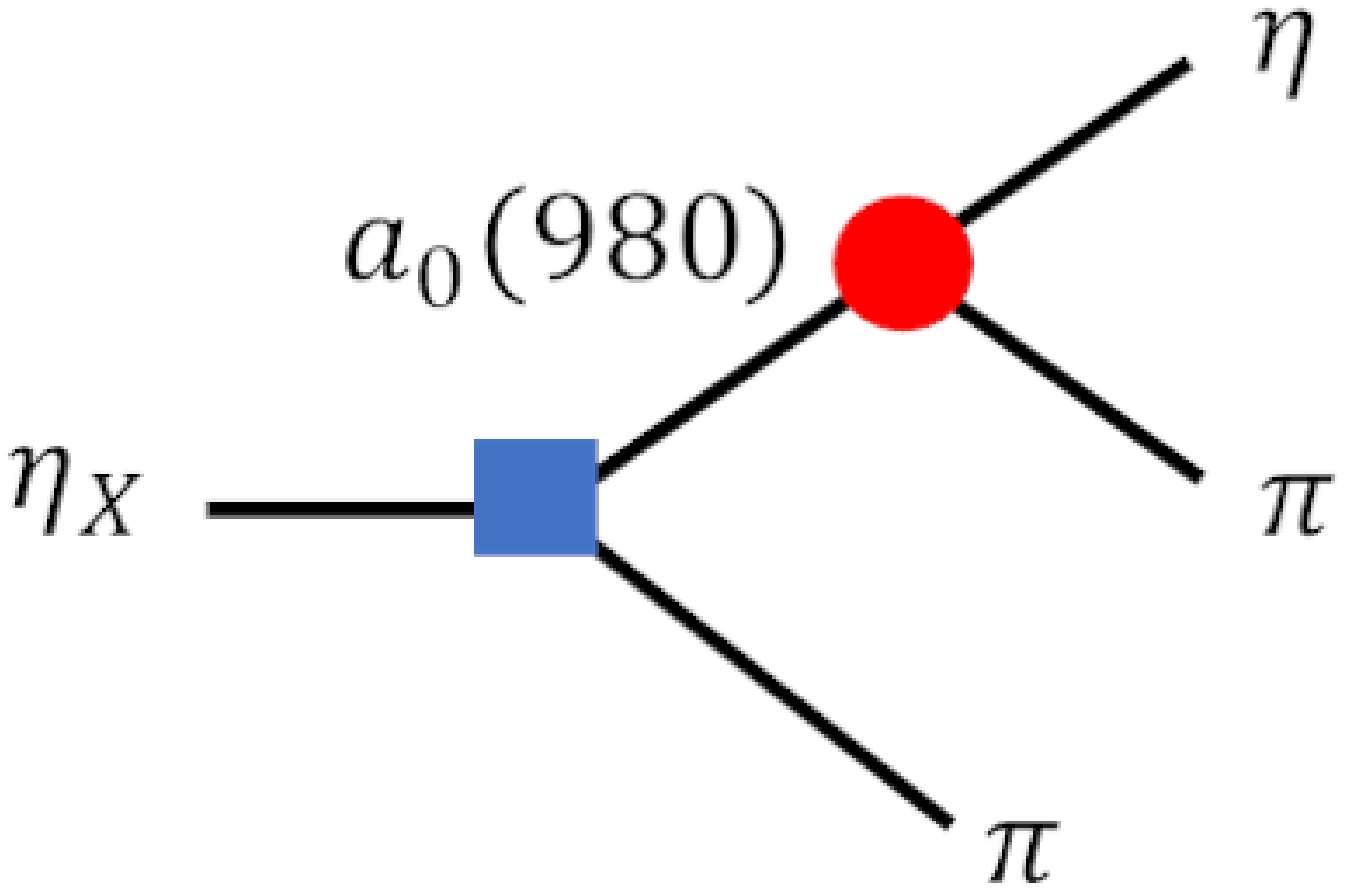}} \qquad 
  \subfigure[]{\includegraphics[width=1.5in]{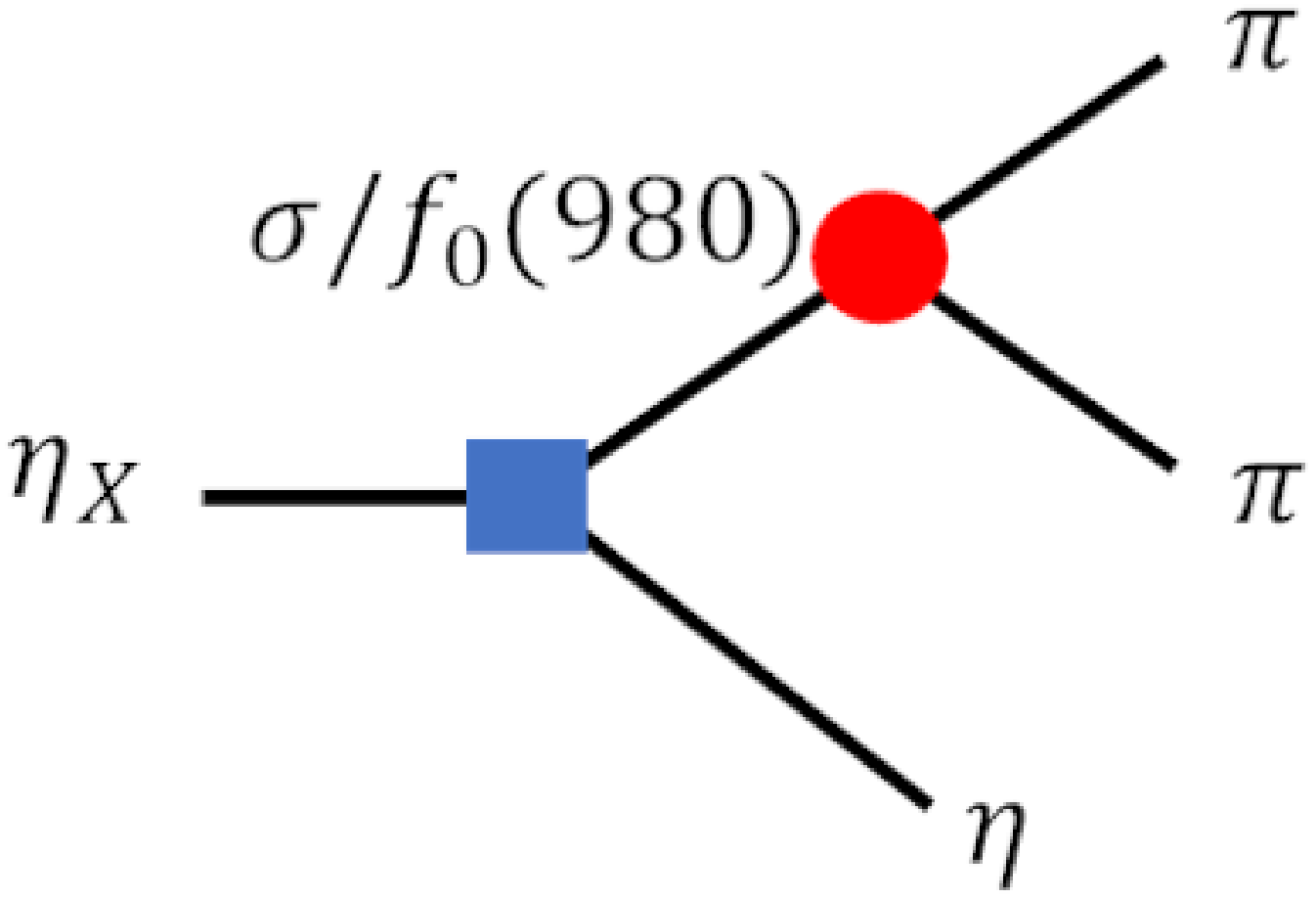}}   \\
  \subfigure[]{\includegraphics[width=1.5in]{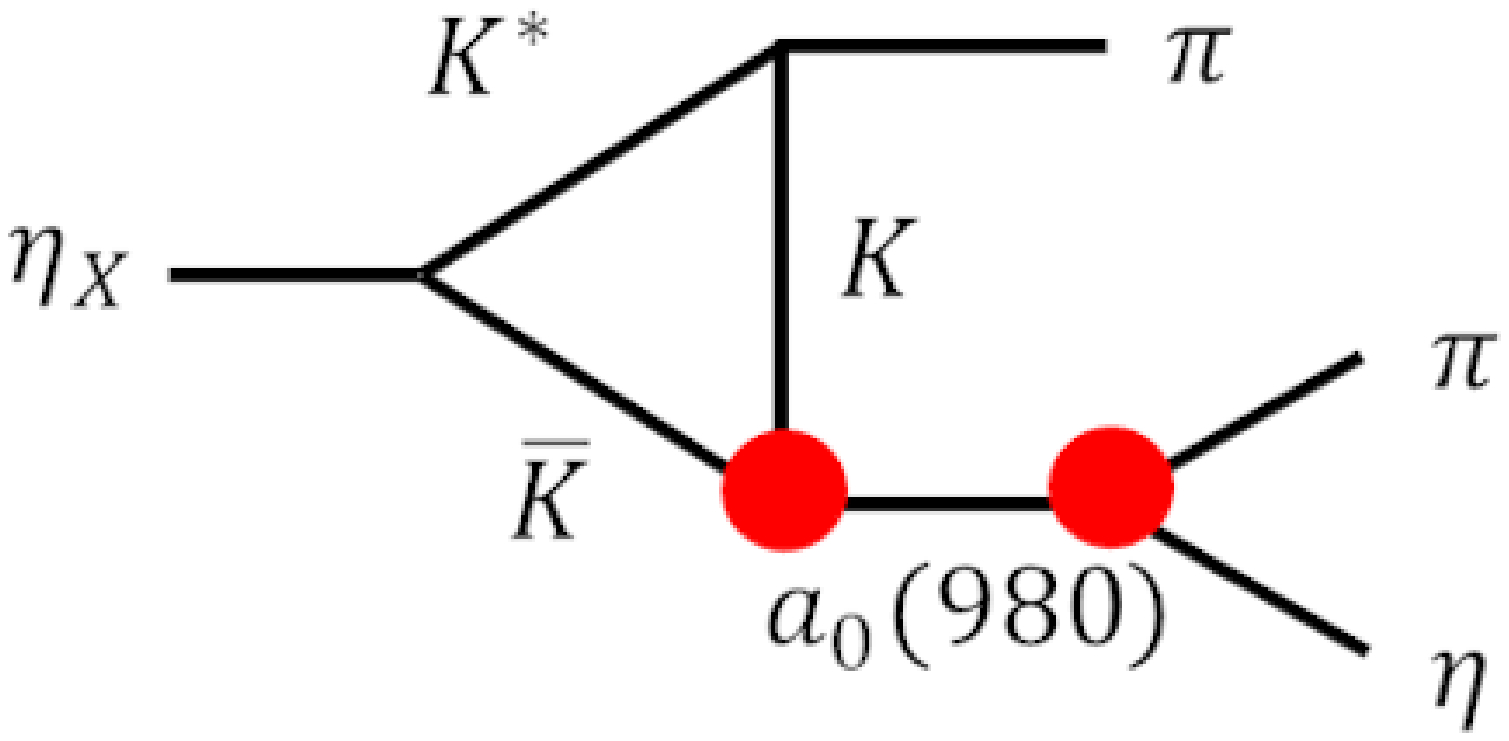}}  \qquad 
  \subfigure[]{\includegraphics[width=1.5in]{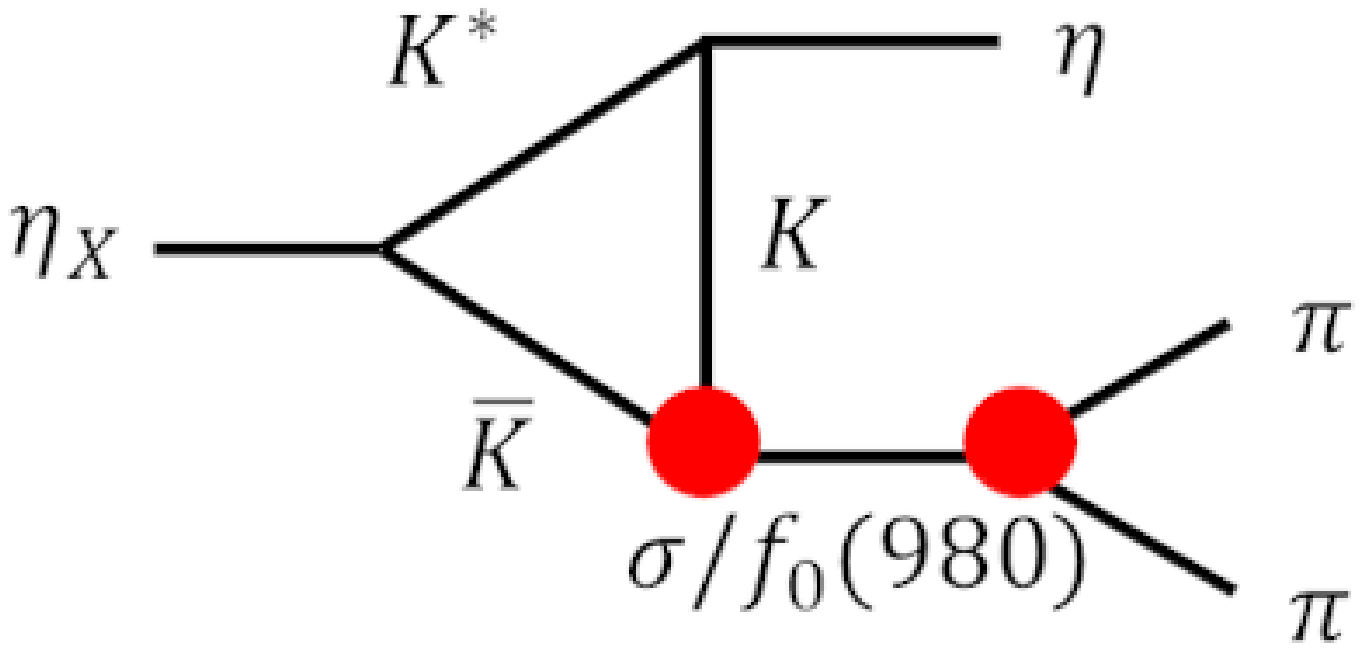}} 
  \caption{Tree level and one-loop diagrams for $\eta_X \to \eta \pi \pi $. The blue solid squares indicate the $XSP$ couplings and the red solid circles indicate the $SPP$ couplings.} \label{fig:etapipidiagram}
\end{figure}

\begin{figure}
     \centering 
 \includegraphics[width=1.6in]{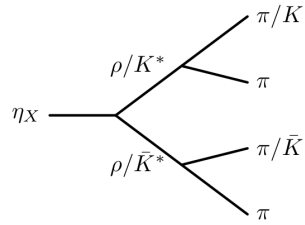}
 \caption{Diagram of four-body decay channels.}\label{fig:fourbody}
\end{figure}

Based on the first radial excitation scenario and with the association of the TS mechanism due to the intermediate $\bar{K}K^*+c.c.$ rescatterings, we plot all possible tree-level and one-loop diagrams that can contribute to the dominant $K \bar{K} \pi$ and $\eta \pi \pi$ channels in Figs.~\ref{fig:kkpidiagram} and~\ref{fig:etapipidiagram}, respectively. In Fig.~\ref{fig:fourbody} we also include the four-body decays of $\eta_X$. We mention in advance that although the four-body decays of $\eta_X$ are relatively small, they may shed some insights into the nature of $\eta_X$ given sufficiently high statistics. 

The tree-level diagrams include the direct couplings of $\eta_X\to \bar{K}K^*+c.c.$, $\kappa \bar{K} +c.c. $, $a_0(980)\pi$, $\sigma \eta$ and $f_0\eta$, while the one-loop diagrams only contain the loop transitions via the intermediate $\bar{K}K^*+c.c.$.

We use effective Lagrangians to describe all the strong vertices appearing in the decay transitions. The hadronic couplings can be arranged by the SU(3) symmetry and then their relative strengths and phases are fixed, especially for those couplings between genuine $q\bar{q}$ states.  There are two types of hadronic coupling vertices involved in this study, i.e., $VPP$ and $SPP$, for which the corresponding effective Lagrangians are as follows:
\begin{eqnarray}
  \mathcal{L}_{VPP}&=&i g_{VPP} \text{Tr}[(P \partial_{\mu}P - \partial_{\mu}P P)V^{\mu}] , \label{Lagrangianvpp} \\
   \mathcal{L}_{SPP}&=& g_{SPP} \text{Tr}[S P P], \label{Lagrangianspp}
\end{eqnarray}
where $S$, $P$ and $V$ stand for the scalar, pseudoscalar and vector fields, respectively, in the flavor SU(3) multiplets, and they have the following forms:
\begin{equation}\label{su3-scalar}
S=\begin{pmatrix}
  \frac{ \sigma + a_0(980)}{\sqrt{2}} &a_0^+                               & \kappa^+ \\
    a_0^-                           &  \frac{ \sigma-a_0(980)}{\sqrt{2}} & \kappa^0 \\
    \kappa^-                        &   \bar{\kappa}^0                   &f_0(980)
\end{pmatrix},
\end{equation}
\begin{equation}\label{su3-pseudo}
  P=\left(
    \begin{array}{ccc}
      \frac{\sin\alpha_P \eta'+ \cos \alpha_P\eta+\pi^0}{\sqrt{2}} & \pi^{+}& K^{+}\\
      \pi^{-} & \frac{ \sin\alpha_P \eta'+ \cos \alpha_P\eta-\pi^0}{\sqrt{2}}& K^{0} \\
      K^{-} & \bar{K^{0}}& \cos\alpha_P \eta'-\sin\alpha_P \eta \\
    \end{array}
\right),
\end{equation}
and 
\begin{equation}\label{su3-vector}
  V=\left(
    \begin{array}{ccc}
      \frac{\omega+\rho^{0}}{\sqrt{2}} & \rho^{+} & {K^{*}}^+ \\
      \rho^{-} &  \frac{\omega-\rho^{0}}{\sqrt{2}} & K^{*0} \\
      K^{*-}& \bar{K}^{*0} & \phi \\
    \end{array}
  \right) ,
\end{equation}
where the ideal mixing are adopted between $\sigma$ and $f_0(980)$, and between $\omega$ and $\phi$.

By expanding Eq.~(\ref{Lagrangianvpp}) we obtain the specific expression of the effective Lagrangian for the $K^* K \pi$ vertex:
\begin{eqnarray}
  \mathcal{L}_{K^{*0} K^0 \pi^0 }=i \frac{g_{VPP}}{\sqrt{2}} ( \pi^0 \partial_{\mu}K^0 -K^0 \partial_{\mu} \pi^0)(K^{*0})^\mu,
\end{eqnarray} 
where the coupling $g_{VPP}$ can be determined by the experimental data for $K^*\to K\pi$.

Assuming that the first radial excitation nonet has the same form as Eq.~(\ref{su3-pseudo}) we describe the mixing between the two isoscalars as follows~\cite{Wu:2011yx,Wu:2012pg,Du:2019idk,Cheng:2021nal}: 
\begin{eqnarray}
       \eta(1295)&=&\cos \alpha_P n\bar{n}-\sin \alpha_P s\bar{s},\\
       \eta(1405)&=&\sin \alpha_P n\bar{n}+\cos \alpha_P s\bar{s},
\end{eqnarray}
where $n\bar{n} \equiv (u\bar{u}+d \bar{d})/ \sqrt{2}$, and $\alpha_P \equiv \arctan \sqrt{2}+\theta_p$ with $\theta_p$ the flavor singlet and octet mixing angle. Although it is still an open question that whether the mixing angle $\alpha_P$ is the same as that for the $\eta-\eta'$ mixing, we adopt the same mixing angle $\alpha_P=42^\circ$ in the calculation. With the mixing angle the strong couplings of $\eta_X$ can in general be expressed as an overall coupling constant multiplied by a factor contributed by the mixing, i.e.,
\begin{eqnarray}\label{1405triangle1}
  \mathcal{L}_{\eta(1405)K^{*0}\bar{K}^0}&=& i g_{\eta(1405)K^{*0} \bar{K}^0}(\bar{K}^0 \partial_\mu \eta(1405)-\eta(1405) \partial_\mu \bar{K}^0 )(K^{*0})^\mu \nonumber\\
  &\equiv & i g_{XVP} (\frac{ \sin \alpha_P}{\sqrt{2}} R -\cos \alpha_P)(\bar{K}^0 \partial_\mu \eta(1405)-\eta(1405) \partial_\mu \bar{K}^0 )(K^{*0})^\mu \ ,
\end{eqnarray}
and 
\begin{eqnarray}\label{1295triangle1}
 \mathcal{L}_{\eta(1295)K^{*0} \bar{K}^0 }&=& i g_{\eta(1295)K^{*0} \bar{K}^0}(\bar{K}^0 \partial_\mu \eta(1295)-\eta(1295) \partial_\mu \bar{K}^0 )(K^{*0})^\mu \nonumber\\
 &\equiv & i g_{XVP} (\frac{\cos \alpha_P}{\sqrt{2}} R + \sin\alpha_P)(\bar{K}^0 \partial_\mu \eta(1295)-\eta(1295) \partial_\mu \bar{K}^0 )(K^{*0})^\mu \ ,
\end{eqnarray}
where $g_{XVP}$ is the overall coupling between a radial excitation pseudoscalar $(q\bar{q})_{0^{-+}}$ and $VP$; $R$ is the SU(3) flavor-symmetry-breaking factor which is also utilized in Ref.~\cite{Cheng:2021nal}.
Similar relations are also present in the $S$-wave couplings $g_{\eta(1295)SP}$ and $g_{\eta(1405)SP}$ where an overall coupling $g_{XSP}$ between a radial excitation pseudoscalar $(q\bar{q})_{0^{-+}}$ and the scalar-pseudoscalar pair $SP$ can be defined.

One notices that only the $\bar{K} K^*+c.c.$ rescattering loops are included in the decay transitions in Figs.~\ref{fig:kkpidiagram} and ~\ref{fig:etapipidiagram}.
This is based on two observations. Firstly, the intermediate $\bar{K} K^*+c.c.$ channel plays a crucial role due to the TS mechanism. 
In contrast, we adopt the experimental constraints on the couplings for the initial pseudoscalar $\eta_X\to SP$ which means that the rescattering effects for other channels have been absorbed into the effective couplings defined for the tree diagrams. 
Secondly, as shown in Ref.~\cite{Du:2022nno}, the vertex corrections from the $\bar{K} K^*+c.c.$ TS mechanism to the leading order bare coupling between $\eta_X$ and $K^*\bar{K}$ is small. 
It means that we can reliably determine the $\eta_X$ couplings to $\bar{K} K^*+c.c.$ to the order of one loop and then apply them to the triangle transitions for other channels as illustrated in Figs.~\ref{fig:kkpidiagram} and~\ref{fig:etapipidiagram}.

As the consequence of the small higher-order corrections to the $\eta_X K^* \bar{K}$ couplings, one recognizes that the SU(3) flavor relation should hold between couplings $g_{\eta(1295) K^* \bar{K}}$ and $g_{\eta(1405) K^* \bar{K}}$. It thus allows us to assume that the coupling $g_{XVP}$ has the same sign as the ground-state couplings $g_{VPP}$, which is consistent with the assumption made in Ref.~\cite{Cheng:2021nal}.
Namely, we take the same sign for the overall coupling constants $g_{VPP}$ and $g_{XVP}$ and they are defined as positive and real numbers, and then the signs for the other $VPP$ and $XVP$ couplings can be fixed.
 We extract the overall coupling $g_{VPP}$ from the partial decay width of $K^*\to K \pi$~\cite{Workman:2022ynf}, and the strength of $g_{XVP}$ is left to be a free parameter to be fitted by the $\eta(1405/1475)$ data. In Tab.~\ref{tab:vppxvpcouplings} we have listed the expressions and values for the $VPP$ and $XVP$ couplings under the SU(3) flavor symmetry.

\begin{table}
  \centering
  \caption{Strong couplings for the $VPP$ and $XVP$ vertices. The value of $g_{\eta(1405)K^*\bar{K}}$ is adopted from fit Scheme-I.}
  \begin{tabular}{c|c|c}
    \hline \hline
    Coupling const.  & Expression   &  Values \\
    \hline
    $g_{VPP}$                           &     ...                                                  & $4.53$   \\
    $g_{K^{*0} K^0 \pi^0}$              &     $g_{VPP}/\sqrt{2}$                                   &  $3.21$   \\
     $g_{K^{*0} K^0 \eta }$             &     $-(\cos\alpha_P/\sqrt{2}+\sin\alpha_P)g_{VPP}$       &   $-5.41$  \\
    \hline
     $g_{XVP}$                          &     ...                                                  & $8.51 \pm 0.31 $  \\
     $g_{\eta(1405) K^{*0} \bar{K}^0}$  &     $(R\sin\alpha_P /\sqrt{2} -\cos\alpha_P)g_{XVP}$     & $-3.12\pm 0.11$     \\
     $g_{\eta(1295) K^{*0}\bar{K}^0}$   &     $(R\cos\alpha_P /\sqrt{2}+ \sin \alpha_P)g_{XVP}$    & $9.26 \pm 0.33$  \\
     \hline \hline
  \end{tabular}\label{tab:vppxvpcouplings}
\end{table}

\subsubsection{Tree-level amplitudes}

With the effective Lagrangians defined earlier, the amplitudes for the tree diagrams can be obtained. 
For the $K\bar{K}\pi$ decay channel, the amplitude for Fig.~\ref{fig:kkpidiagram}(a) for the $K^0 \bar{K}^0 \pi$ channel can be expressed as
\begin{eqnarray}
  i \mathcal{M}_{K^*\bar{K}}=&-i&\biggl(g_{\eta_X K^{*0} \bar{K}} g_{K^{*0}K^0 \pi }\frac{(2p_{X}-p_{ab})_\mu (g^{\mu \nu}-\frac{p_{ab}^\mu p_{ab}^\nu}{p_{ab}^2})(p_{ab}-2p_b)_{\nu}}{p_{ab}^2-m_{K^*}^2+i m_{K^*} \Gamma_{K^*}}  \\ \nonumber
  &+& g_{\eta_X \bar{K}^{*0} K} g_{\bar{K}^{*0}\bar{K}^0\pi} \frac{(2p_{X}-p_{bc})_\mu (g^{\mu \nu}-\frac{p_{bc}^\mu p_{bc}^\nu}{p_{bc}^2})(p_{bc}-2p_b)_{\nu}}{p_{bc}^2-m_{K^*}^2+i m_{K^*} \Gamma_{K^*}} \biggr),
\end{eqnarray}
with $p_{ab}\equiv (p_a + p_b)$ and $p_{bc}\equiv (p_b+p_c)$. To denote the four-vector momenta of the particles, we adopt the following notations for the kinematic variables: $p_X (\eta_X)$, $p_a(K)$, $p_b (\pi )$, $p_c(\bar{K})$ denote the four-vector momenta in the $K\bar{K}\pi$ channel and  $p_a(\pi^0/\pi^+)$, $p_b(\eta )$, $p_c(\pi^0 / \pi^-)$ for those in the $\eta \pi \pi $ channel. We also define the following quantities:
\begin{eqnarray*}
  p_{ac}=p_a+p_c,  \qquad
  s_{ab}=p_{ab}^2, \qquad
  s_{bc}=p_{bc}^2,  \qquad
  s_{ac}=p_{ac}^2,
\end{eqnarray*}
and we will continue to use this notation in the subsequent calculations.
The amplitudes of Fig.~\ref{fig:kkpidiagram}(b) and (c) for $\eta_X\to K^0 \bar{K}^0 \pi$ can be expressed, respectively, as
\begin{eqnarray}
  i \mathcal{M}_{\kappa \bar{K}}=- \biggl(g_{\eta_X \kappa \bar{K}} g_{\kappa K \pi} G_{\kappa} (s_{ab})+  g_{\eta_X \bar{\kappa} K} g_{ \bar{\kappa}\bar{K} \pi} G_{\kappa}(s_{bc})\biggr) \ ,
\end{eqnarray}
and 
\begin{eqnarray}
  i \mathcal{M}_{a_0 \pi }=- g_{\eta_{X}a_0 \pi }g_{a_0 K^0 \bar{K}^0} G_{a_0}(s_{ac}) \ ,
\end{eqnarray}
where $G_{\kappa}(s_{bc})$ and $G_{a_0}(s_{ac})$ are the propagators of $\kappa$ and $a_0$, respectively. 

For the broad scalars, we use the commonly adopted energy-dependent parametrization form in order to simplify the numerical calculations encountered in later loop calculations:
\begin{eqnarray}\label{Eq:EDWidth1}
  G_{S}(s)=\frac{i }{s-m^2_{S}+ i \sqrt{s} \Gamma_{S}(s)},
\end{eqnarray}
with
\begin{eqnarray}\label{Eq:EDWidth2}
  \Gamma_S(s)= \frac{g^2_{S} \cdot k_{S}}{8 \pi s},
\end{eqnarray}
where $k_{S}$ is the magnitude of the final-state pseudoscalar momentum in the center of mass (c.m.) system of the scalar; $m_S$ is the physical mass of the scalar $S$, and the coupling constant $g_{S}$ is defined the partial decay width:
\begin{eqnarray}\label{Eq:EDWidth3}
  \Gamma_S\equiv \Gamma_S(m^2_S)=\frac{g^2_{S} \cdot k_{S}}{8 \pi m^2_S}.
\end{eqnarray}
Note that the decay widths are not sensitive to these parameters. Therefore, we only adopt the best fitted Breit-Wigner masses and widths from experiment and do not consider the uncertainties raised by their errors. For $\kappa$ and $\sigma$, the PDG values are adopted, i.e., $m_{\kappa}=0.826$ GeV, $\Gamma_\kappa=0.449$ GeV, $m_\sigma=0.513$ GeV and $\Gamma_\sigma=0.335$ GeV~\cite{Workman:2022ynf}. The extracted $SPP$ couplings are listed in Tab.~\ref{tab:sppcouplings}.

\begin{table}
  \centering
  \caption{Strong couplings for the $SPP$ vertices.}
  \begin{tabular}{c|c|c}
    \hline \hline
    Parameters  & Expressions in SU(3) symmetry   &   Values  (GeV) \\
    \hline
    $g_{\sigma \pi^0 \pi^0} $    &  $\sqrt{2} g_{SPP}$                                       &  $2.60$ \\
    $g_{\sigma K^0 \bar{K}^0}$   &  $g_{SPP} / \sqrt{2}$                                     &  $1.30$  \\                                      
    \hline
    $g_{a_0\eta\pi^0 }$          &  $ \sqrt{2} g_{SPP}  \cos \alpha_P$                         &  $3.02 \pm 0.35 $  ~\cite{KLOE:2002kzf}   \\
    $g_{a_0 K^0 \bar{K}^0}$      &  $-g_{SPP}/ \sqrt{2}$                                     &  $-2.24 \pm 0.11$  ~\cite{KLOE:2002kzf}   \\
   \hline
    $g_{f_0 \pi^0 \pi^0}$        &  $ 0 $                                                    &  $2.96\pm 0.12$ ~\cite{KLOE:2002deh}     \\
    $g_{f_0 K^0 \bar{K}^0}$      &  $ g_{SPP} $                                              &  $5.92 \pm 0.13$ ~\cite{KLOE:2002deh}   \\
    \hline 
    $g_{\kappa^0 K^0 \pi ^0}$    &  $-g_{SPP}/\sqrt{2}$                                      &  $-3.28$                    \\
   \hline \hline
  \end{tabular}\label{tab:sppcouplings}
\end{table}


Similarly, the tree-level amplitudes for the $\eta \pi^0 \pi^0$ decay channel (Figs.~\ref{fig:etapipidiagram}(a) and (b)) can be expressed as the following forms:
\begin{eqnarray} \label{Eq:2(a)}
    i \mathcal{M}'_{a_0 \pi}=- \frac{1}{\sqrt{2}} g_{\eta_X a_0 \pi} g_{a_0 \eta \pi} \bigg(G_{a_0}(s_{ab})+G_{a_0}(s_{bc})\bigg),
\end{eqnarray}
and
\begin{eqnarray} \label{Eq:2(b)1}
   i \mathcal{M}'_{(\sigma/f_0) \eta}&=&-\frac{1}{\sqrt{2}} g_{\eta_X (\sigma/f_0) \eta} g_{(\sigma/f_0) \pi \pi } G_{(\sigma/f_0)}(s_{ac}) \ .
\end{eqnarray}

Note that the coupling $g_{\eta(1405) a_0 \pi}$ extracted in Ref.~\cite{Du:2019idk} is much smaller than the ordinary hadronic couplings. This is due to the destructive interference between the tree-level bare coupling and triangle amplitude as required by the relative B.R. fractions between $\Gamma(\eta(1405)\to a_0 \pi \to \eta \pi^0 \pi^0 )$ and $\Gamma((\eta(1405) \to K \bar{K}\pi)$. In this study the destructive interference is confirmed by the relative phase angle $\phi_{XSP}$ introduced between the tree-level and triangle loop amplitudes. The $XSP$ couplings are listed in Tab.~\ref{tab:xspcouplings}.

\subsubsection{Triangle loop amplitudes}

\begin{figure}
  \centering
  \includegraphics[width=1.7in]{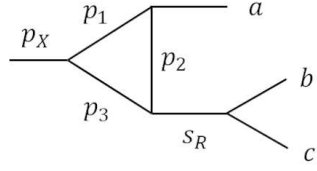}
  \caption{Conventions for the kinematics of the loop diagrams. $s_R$ is the invariant mass square of the intermediate resonance which decays to the final states, therefore, $s_R$ could possibly be $s_{ab}$, $s_{ac}$ or $s_{bc}$. }\label{fig:loopkinematics}
\end{figure}

For the loop transitions illustrated in Figs.~\ref{fig:kkpidiagram} and \ref{fig:etapipidiagram}, the kinematic variables of the loop diagrams are denoted in Fig.~\ref{fig:loopkinematics}. The mass and momentum of the $i$-th internal particle are labelled as $m_i$ and $p_i$. To cut off the UV divergence in the loop integrals, we include a commonly adopted form factor to regularize the integrand,
\begin{eqnarray}
  \mathcal{F}(\mathbf{p}^2_i)=\prod_i \exp \bigg(-\frac{\mathbf{p}^2_i}{\Lambda^2} \bigg),
\end{eqnarray}
where $\Lambda$ is the cutoff energy and its typical value is around the $\rho$ mass; $\mathbf{p}_i$ is the three-vector momentum of the $i$-th particle in the loops in the c.m. frame of the initial particle.

Hence, without considering the couplings, the general form of the triangle loop amplitude $\mathcal{I}$ can be expressed as the following:
\begin{eqnarray}
   \mathcal{I}_{VP(S)}(s_R)=i \int \frac{d^4 p_1}{(2\pi)^4}  \frac{(2p_X -p_1)_\mu (-g^{\mu\nu}+\frac{p^\mu_1 p^\nu_1}{p^2_1})(p_1 -2 p')_\nu}{(p^2_1-m^2_1)(p_2^2-m^2_2)(p_3^2-m^2_3)} G_S(s_R)\mathcal{F} (\mathbf{p}^2_i) ,
\end{eqnarray}
where the subscript $VP(S)$ denotes the intermediate $K^* \bar{K} $ rescattering into a scalar ($S$) and a pseudoscalar ($P$) followed by the scalar decays into two pseudoscalars. Note that $p'$ is the momentum of the external particle which is not from the scalar meson decay.

The triangle amplitude of $\eta_X \to  K^*\bar{K}(\bar{\kappa})\to K^0 \bar{K}^0 \pi^0$ is written as:
\begin{eqnarray}
i \mathcal{M}_{ K^*\bar{K}(\bar{\kappa})} = \hat{g}_\kappa [(\mathcal{I}^N(s_{ab})+\mathcal{I}^N(s_{cb}))+ 2  (\mathcal{I}^C(s_{ab})+\mathcal{I}^C(s_{cb}))],
\end{eqnarray}
where the superscripts $C$ and $N$ in the loop function $\mathcal{I}$ indicate the charged and neutral loop, respectively; $\hat{g}_\kappa$ represents the product of the combined vertex couplings in the $K^*\bar{K}(\bar{\kappa})$ loop. 
The subscript $VP(S)$ of $\mathcal{I}$ has been omitted for brevity. We adopt the notation $\mathcal{I}$ for the loop function in the loop amplitudes. Its different expressions in different triangle loops are implied.

For the process of  $\eta_X \to  K^*\bar{K}(a_0)\to K^0 \bar{K}^0 \pi^0$ in Fig.~\ref{fig:kkpidiagram}(e), the amplitude is written as:
\begin{eqnarray}
  i \mathcal{M}_{ K^*\bar{K}(a_0)}= 2 \hat{g}_{a_0} [ \mathcal{I}^C(s_{ac})+ \mathcal{I}^N(s_{ac})],
\end{eqnarray}
where $\hat{g}_{a_0}$ represents the product of the combined vertex couplings in the $K^*\bar{K}(a_0)\to K^0 \bar{K}^0 \pi^0$ loop.
For the processes of $\eta_X \to K^*\bar{K} (a_0) \to \eta \pi^0 \pi^0 $ in Fig.~\ref{fig:etapipidiagram}(c), the amplitude is written as
\begin{eqnarray}
  i \mathcal{M'}_{ K^*\bar{K}(a_0)}= \sqrt{2}\hat{g}_{a_0}^\prime [\mathcal{I}^N(s_{ab})+ \mathcal{I}^N(s_{bc})+\mathcal{I}^C(s_{ab})+ \mathcal{I}^C(s_{bc})] , 
\end{eqnarray}
where $\hat{g}_{a_0}^\prime$ represents the product of the combined vertex couplings in the $K^*\bar{K}(a_0)\to \eta \pi^0 \pi^0$ loop.
For the processes of $\eta_X \to K^*\bar{K} (\sigma) \to \eta \pi^0 \pi^0 $ and $\eta_X \to K^*\bar{K} (f_0) \to \eta \pi^0 \pi^0 $ in Fig.~\ref{fig:etapipidiagram}(d), the amplitudes are written respectively as
\begin{eqnarray}
i \mathcal{M'}_{ K^*\bar{K}(\sigma)}= \sqrt{2} \hat{g}_\sigma [\mathcal{I}^C(s_{ac})+ \mathcal{I}^N(s_{ac})]
\end{eqnarray}
and 
\begin{eqnarray}
  i \mathcal{M'}_{ K^*\bar{K}(f_0)}= \sqrt{2} \hat{g}_{f_0} [\mathcal{I}^C(s_{ac})+ \mathcal{I}^N(s_{ac})],
  \end{eqnarray}
where $\hat{g}_\sigma$ and $\hat{g}_{f_0}$ are the products of the combined vertex couplings in the $K^*\bar{K}(\sigma)$ and $K^*\bar{K}(f_0)$ loops, respectively.

Finally, we collect the amplitudes and express the total amplitudes for each channel as follows: With
\begin{eqnarray}
  \mathcal{M}_{K^*\bar{K}\to K^0\bar{K}^0 \pi^0} &=&\mathcal{M}_{K^*\bar{K}}, \\
  \mathcal{M}_{\kappa \bar{K}\to  K^0\bar{K}^0 \pi^0} &=& \mathcal{M}_{\kappa \bar{K}}+\mathcal{M}_{K^*\bar{K}(\kappa)}, \\
  \mathcal{M}_{a_0 \pi \to  K^0\bar{K}^0 \pi^0} &=& \mathcal{M}_{a_0 \pi}+\mathcal{M}_{K^*\bar{K}(a_0)}, 
\end{eqnarray}
we obtain
\begin{eqnarray}
  \mathcal{M}_{ K^0\bar{K}^0 \pi^0}&=& \mathcal{M}_{K^*\bar{K}\to K^0\bar{K}^0 \pi^0}+\mathcal{M}_{\kappa \bar{K}\to  K^0\bar{K}^0 \pi^0}+ \mathcal{M}_{a_0 \pi \to  K^0\bar{K}^0 \pi^0}.
\end{eqnarray}
With
\begin{eqnarray}
  \mathcal{M}_{\sigma \eta \to \eta \pi^0 \pi^0}&=&\mathcal{M}'_{\sigma \eta}+\mathcal{M}'_{K^*\bar{K}(\sigma)}, \\
  \mathcal{M}_{f_0 \eta \to  \eta \pi^0 \pi^0} &=& \mathcal{M}'_{f_0 \eta}+\mathcal{M}'_{K^*\bar{K}(f_0)}, \\
  \mathcal{M}_{a_0 \pi \to \eta \pi^0 \pi^0} &=& \mathcal{M}'_{a_0 \pi}+\mathcal{M}'_{K^*\bar{K}(a_0)}, 
\end{eqnarray}
we obtain
\begin{eqnarray}
  \mathcal{M}_{ \eta \pi^0 \pi^0}&=& \mathcal{M}_{\sigma \eta \to \eta \pi^0 \pi^0}+\mathcal{M}_{f_0 \eta \to  \eta \pi^0 \pi^0}+ \mathcal{M}_{a_0 \pi \to \eta \pi^0 \pi^0}.
\end{eqnarray}

Consequently, the partial widths of the processes to $K\bar{K} \pi $ and $\eta \pi \pi $ channels are respectively,  
\begin{eqnarray}\label{eq:partialwidthkkpi}
      \Gamma_{\eta_X \to K\bar{K}\pi} = 6 \Gamma_{K^0 \bar{K}^0 \pi^0}=\frac{6}{2\sqrt{s}} \int d \Phi_{K^0 \bar{K}^0 \pi} |\mathcal{M}_{ K^0\bar{K}^0 \pi^0}|^2
\end{eqnarray}
and
\begin{eqnarray}\label{eq:partialwidthetapipi}
  \Gamma_{\eta_X \to \eta \pi \pi } = 3 \Gamma_{ \eta \pi^0 \pi^0}=\frac{3}{2\sqrt{s}} \int d \Phi_{\eta \pi^0 \pi^0} |\mathcal{M}_{\eta \pi^0 \pi^0}|^2,
\end{eqnarray}
where $\Phi_{abc}$ is the phase space of $\eta_X \to abc$.
To calculate the partial widths of the decay to $K\bar{K}\pi$ ($\eta \pi \pi $) channel from a given intermediate states, one can just replace the total amplitude in Eq.~(\ref{eq:partialwidthkkpi}) (Eq.~(\ref{eq:partialwidthetapipi})) with the amplitudes of the corresponding intermediate states.

\subsubsection{Parameter constraint and fitting schemes}

\begin{table}
  \centering
  \caption{Strong couplings for the $XSP$ vertices.}
  \begin{tabular}{c|c|c}
    \hline \hline
    Parameters  & Expressions in SU(3) symmetry   &  Absolute values (GeV) \\
    \hline
    $g_{XSP}$                             &           ...                                        &  $0.45 \pm 0.06$ \\
  \hline
    $g_{\eta(1405) \sigma \eta}$          &   $\sqrt{2} g_{XSP} \cos\alpha_P \sin \alpha_P$      &  $0.32\pm0.04$ \\
    $g_{\eta(1295) \sigma \eta}$          &   $\sqrt{2} g_{XSP} \cos^2\alpha_P $                 &  $0.35\pm0.05$  \\
  \hline
    $g_{\eta(1405) f_0 \eta }$            &   $-2 g_{XSP} \cos\alpha_P \sin\alpha_P$             &  $-0.45\pm0.06$ \\
    $g_{\eta(1295) f_0 \eta}$             &   $2 g_{XSP} \sin^2 \alpha_P $                       &  $0.40\pm0.05$    \\
  \hline
    $g_{\eta(1405) a_0 \pi^0}$            &  $ \sqrt{2} g_{XSP}\sin\alpha_P$                      &  $0.43\pm0.06$ \\
    $g_{\eta(1295) a_0 \pi^0 }$           &   $\sqrt{2} g_{XSP} \cos \alpha_P$                    &  $0.47\pm0.06$   \\
  \hline
    $g_{\eta(1405) \kappa^0 \bar{K}^0}$   & $g_{XSP} (\sin \alpha_P/\sqrt{2}+\cos\alpha_P)$      &  $0.50\pm0.07$                             \\
    $g_{\eta(1295) \kappa^0 \bar{K}^0}$   & $g_{XSP} (\cos \alpha_P/\sqrt{2}-\sin\alpha_P)$      &  $-0.11\pm0.02$                           \\
  \hline \hline
  \end{tabular}\label{tab:xspcouplings}
\end{table}

In order to keep self-consistent with our theoretical approach based on the one-state scenario for $\eta(1405)$ and $\eta(1475)$ we adopt the data for the signals around 1.4 GeV and treat them as from the same state. Namely, if two states were adopted in the analysis, and only one state is seen, we treat it as from a single state. As the consequence, all the $0^{-+}$ signals around 1.4 GeV from the charmonium decays are treated as from a single state $\eta(1405)$.
Considering the PDG data are not self-consistent, we choose the data as input in our fitting via the rules as following: data in the relatively recent are chosen; data analyzed by the one resonance PWA (partial wave analysis) are chosen and the two resonance PWA data are excluded.

Because $\kappa$ and $\sigma$ are of $S$ wave and have broad line shapes, it is difficult to isolate their signals from the background contributions in experiment. 
Therefore, there might be large uncertainties in the $\kappa$ and $\sigma$ data.
We perform two fit schemes which adopt different data sets as input. In Scheme-I we only adopt the following data as input regarding their relatively high statistics:
\begin{itemize}
  \item  $\Gamma(\eta \pi \pi )/\Gamma(K\bar{K}\pi)= 0.16 \pm 0.04$
  \item  $\Gamma(K^*\bar{K}\to K\bar{K} \pi ) / \Gamma(K \bar{K} \pi )= 0.70 \pm 0.05 $ 
  \item  $\Gamma(a_0(980)\pi \to \eta \pi \pi ) / \Gamma(\eta \pi \pi )=0.56 \pm 0.07 $
  \item  $\Gamma_{\eta(1405)}=50.1 \pm 2.6$ MeV
\end{itemize}
Here we use the largest PDG value of $\Gamma(a_0(980)\pi \to \eta \pi \pi ) / \Gamma(\eta \pi \pi )=0.56 \pm 0.07$~\cite{Workman:2022ynf}. 

In Scheme-II, in addition to the experimental data adopted in Scheme-I, we also try to fit all the other B.R. fractions plus the total width of $\eta(1405)$  which are listed in Tab.~\ref{Tab:data1405}. One can see that the B.R. fractions of $\Gamma(K^* \bar{K}\to K\bar{K}\pi)/\Gamma(\kappa \bar{K} \to K \bar{K} \pi)$ and $\Gamma(a_0(980)\pi \to \eta \pi \pi) / \Gamma(\eta (\pi \pi )_{S-\text{wave}})$ have different values from different measurements and they are not in agreement with each other. We choose the following data in the Scheme-II fit:
\begin{itemize}
  \item $\Gamma(K^* \bar{K}\to K\bar{K}\pi)/\Gamma(\kappa \bar{K} \to K \bar{K} \pi) =5.4 \pm 1.3$
  \item  $\Gamma(a_0(980)\pi \to \eta \pi \pi) / \Gamma(\eta (\pi \pi )_{S-\text{wave}})=0.91\pm 0.12$ \ .
\end{itemize}
We adopt the cut off energy $\Lambda=0.8$ GeV in the loop amplitudes and the $\eta(1405)$ mass as $1.42$ GeV in the fits. 

\subsection{Four-pseudoscalar decay mechanisms}

Besides the major three-body decay channels, there are possible four-body channels via the decay processes as $\eta_X \to V V \to 4 P$, where $V$ and $P$ denotes the vectors and pseudoscalars as we addressed above. 
The processes of such type include $\rho \rho \to 4 \pi$, $K^* \bar{K}^* \to K \bar{K} \pi \pi $.
We calculate these two transitions to study whether such four-body decays will have considerable contributions to the widths of $\eta_X$.
The couplings of $\eta_X\to V V$ have been learned in Ref.~\cite{Cheng:2021nal}, and the coupling $g_{\rho \pi \pi }=5.9$ can be extracted from the $\rho\to\pi\pi$ data.  

The amplitude of the tree-level transition $\eta_X \to \rho \rho \to 4 \pi$ has the form as:
\begin{eqnarray}
   i \mathcal{M}=g_{\eta_X \rho\rho} g^2_{\rho\pi\pi}\frac{ \epsilon_{\alpha \beta \mu \nu } (p_1 +p_2)^\alpha (p_3+p_4)^\beta (p_1-p_2)^\mu (p_3-p_4)^\nu }{ (s_{12}-m^2_{\rho}+i \sqrt{s_{12}}\Gamma_{\rho}(s_{12}))(s_{34}-m^2_{\rho}+i \sqrt{s_{34}}\Gamma_{\rho}(s_{34}))},
\end{eqnarray}
where we use notations $p_1$ and $p_2$ to respectively indicate the four-momentum of the two $\pi$ mesons from one intermediate $\rho$ decay, and $p_3$ and $p_4$ are respectively the four-momentum of the other two $\pi$ from the other  intermediate $\rho$ decay.
Likewise, $s_{12}$ is defined as $(p_1+p_2)^2$ and $s_{34}$ as $(p_3+p_4)^2$. Since $\rho$ is a relatively broad resonance, we use an energy-dependent Breit-Wigner formula for its propagator.
The energy-dependent width $\Gamma_{\rho}(s)$ is defined the same form as Eqs.~(\ref{Eq:EDWidth1}), (\ref{Eq:EDWidth2}), and (\ref{Eq:EDWidth3}).
To be specific, there are two different $4 \pi $ states: $\rho^0 \rho^0 \to 2 \pi^+ 2 \pi^-$ and $\rho^+\rho^- \to 2 \pi^0 \pi^+ \pi^- $.
Hence, the partial decay width for the $4 \pi $ channel is:
\begin{eqnarray}
     \Gamma_{4 \pi}= \frac{1}{2\sqrt{s}}\bigg(\frac{1}{4}\int d\Phi_{2 \pi^+ 2 \pi^-} |\mathcal{M}|^2+ \frac{1}{2} \int d\Phi_{2 \pi^0 \pi^+ \pi^-} |\mathcal{M}|^2\bigg),
\end{eqnarray}
The constant $1/2$ and $1/4$ are the symmetry factors on account of the identical particles. $\Phi_{4 \pi }$ is the four-body phase space of the $4\pi$ final states.
  
Similarly, we can write down the tree-level amplitude for $\eta_X \to K^* \bar{K}^* \to K^0\bar{K}^0 \pi^0 \pi^0 $ as following:
\begin{eqnarray}
  i \mathcal{M}=g_{\eta_X K^{*0}\bar{K}^{*0}} g^2_{K^{*0} K^0 \pi^0 }\frac{ \epsilon_{\alpha \beta \mu \nu } (p_1 +p_2)^\alpha (p_3+p_4)^\beta (p_1-p_2)^\mu (p_3-p_4)^\nu }{ (s_{12}-m^2_{K^*}+i m_{K^*} \Gamma_{K^*})(s_{34}-m^2_{K^*}+i m_{\bar{K}^*}\Gamma_{\bar{K}^*})},
\end{eqnarray}
For this case, $p_1$ and $p_2$ indicate the four-momentum of $K$ and $\pi$ that come from the intermediate $K^*$, and $p_3$ and $p_4$ indicate the four-momentum of $\bar{K}$ and $\pi$ that come from the intermediate $\bar{K}^*$.
Because $K^*$ is a relatively narrow state and the energy-independent Breit-Wigner propagator could be a good approximation.
The $K\bar{K} \pi \pi $ channels specifically include: $K^+ K^- \pi^+ \pi^-$, $K^0 \bar{K}^0 \pi^+ \pi^-$,  $K^0 \bar{K}^0 \pi^0 \pi^0$,  $K^+ \bar{K}^- \pi^0 \pi^0$,  $K^0 K^- \pi^+ \pi^0$ and $K^+ \bar{K}^0 \pi^0 \pi^-$.
Note that for the final states including two $\pi^0$ the symmetry factor $1/2$  for the identical particle should be included. 
   
We neglect the mass difference between the charged and neutral states and adopt the same phase space for the above mentioned six $K\bar{K} \pi \pi $ channels. Finally the partial width of the total $K\bar{K}\pi \pi $ channel could be presented by the partial width of the $K^0 \bar{K}^0 \pi^0 \pi^0$ channel multiplied by a factor which could be determined according to the SU(3) flavor symmetry relation among the couplings in these transitions.
The partial decay width for the $K \bar{K} \pi \pi $ channel can be written as 
\begin{eqnarray}
  \Gamma_{K \bar{K}\pi \pi }= \frac{34}{2\sqrt{s}}\bigg(\frac{1}{2}\int d\Phi_{K\bar{K}\pi \pi } |\mathcal{M}|^2 \bigg),
\end{eqnarray}
where $\Phi_{K\bar{K}\pi \pi }$ is the four-body phase space of the $K^0\bar{K}^0\pi^0 \pi^0 $ final states.
The partial decay widths of these two channels are listed in Tab.~\ref{Tab:decaywidths}. 
  
\section{Numerical results and discussions}\label{sec:3}

The best fitted parameters of the two fits are listed in Tab.~\ref{Tab:fittedparas}. The couplings $g_{XVP}$ and $g_{XSP}$ fitted in both Scheme-I and II are consistent with the extracted values in Ref.~\cite{Du:2019idk}. There, values of $10.9$ and $0.76$ are found for $g_{XVP}$ and $g_{XSP}$, respectively.

\begin{table}
  \centering
  \caption{The best fitted parameters from Scheme-I and II.}
  \begin{tabular}{lcccc}
    \hline \hline
                              &  $g_{XVP}$                   &  $g_{XSP}$ (GeV)           & $\phi_{XSP}$                    &  $\phi_{SPP}$  \\ \hline 
  Scheme-I       &       $8.50 \pm 0.31$          &  $0.45 \pm 0.06$           & $0^\circ \pm 13^\circ$          &  $180^\circ \pm 13^\circ$  \\ 
  Scheme-II      &       $6.60 \pm 3.31 $         &  $0.58 \pm 0.69$           &  $36^\circ \pm 280^\circ $    &   $224^\circ \pm 112^\circ $               \\             
\hline \hline 
     \end{tabular}
  \label{Tab:fittedparas}
\end{table}

\subsubsection{Branching ratio fractions}

The available experimental B.R. fractions for $\eta(1405/1475)$ and $\eta(1295)$ are collected in Tabs.~\ref{Tab:data1405} and ~\ref{Tab:data1295}, respectively, among which those adopted as the input data in Scheme-I are highlighted in boldface.  Theoretical calculations of the corresponding observables in Scheme-I and II are also listed in Tabs.~\ref{Tab:data1405} and ~\ref{Tab:data1295}. As shown in Tab.~\ref{Tab:data1405}, the input data in Scheme-I fit are fitted well. In contrast, the other observables collected in this table and calculated with the fitted parameters from Scheme-I seem to have significant discrepancies from the experimental data, especially for the ratios related to the B.R. of $\eta (\pi \pi)_{S-\text{wave}}$, which is considered as the contributions from $\sigma \eta$.

With additional experimental constraint included in the Scheme-II fit, i.e., all the data in Tab.~\ref{Tab:data1405} are adopted as the experimental input, the central values of the fitted B.R. fractions turn out to be comparable with those obtained in Scheme-I. However, much larger uncertainties are found. It implies that the additional input data adopted in Scheme-II are incompatible with other observables in our fitted model. It is also an indication that there could exist some inconsistencies among these measured B.R. fractions and further experimental studies are demanded. 

The predictions for $\eta(1295)$ from the two fitting schemes are listed in Tab.~\ref{Tab:data1295} to compare with the available data. The B.R. fractions of $\Gamma( a_0(980)\pi\to \eta \pi \pi )/ \Gamma( \eta \pi^0 \pi^0 )$  can be well described by the parameters extracted in both Scheme-I and II. In contrast, the theoretical values for the two B.R. fractions, $\Gamma( a_0(980)\pi) / \Gamma( \sigma \eta)$ and $\Gamma( \eta (\pi \pi )_{S-\text{wave}})/ \Gamma( \eta \pi^0 \pi^0)$, turn out to be very different from the measurements. However, this is understandable since these large deviations could be due to the difficulty on treating the isoscalar $(\pi\pi)_{S-wave}$ contributions in the analysis. The broad $\sigma$ as a dynamically generated state in the $\pi\pi$ scatterings cannot be unambiguously separated from the background. Further experimental measurements with the partial wave analysis are necessary for examining the contributions from the isoscalar $(\pi\pi)_{S-wave}$ channel. One notices that the $\eta(1295)$ total width in Scheme-I is significantly larger than the PDG value, while that from Scheme-II turns out to be compatible. It should be noted that the $\eta(1295)$ total width is very sensitive to the parameters for the $S$-wave coupling, i.e., $\eta(1295)\to \sigma \eta$. Namely, it also originates from our poor knowledge about the isoscalar $(\pi\pi)_{S-wave}$ channel.

\subsubsection{Invariant mass spectra}

\begin{figure}
  \centering
  \subfigure[]{\includegraphics[width=3in]{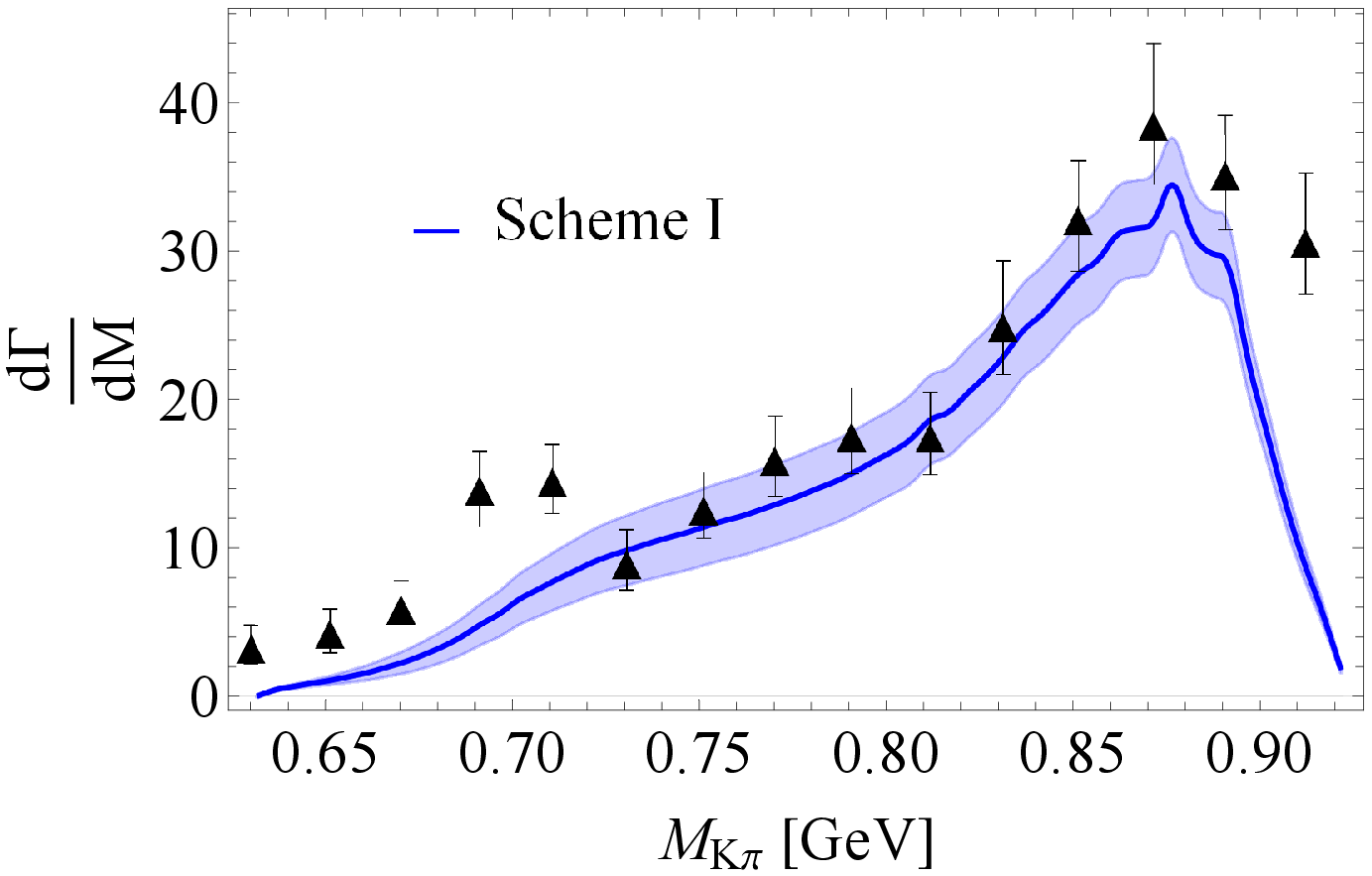}} \qquad \qquad
  \subfigure[]{\includegraphics[width=3in]{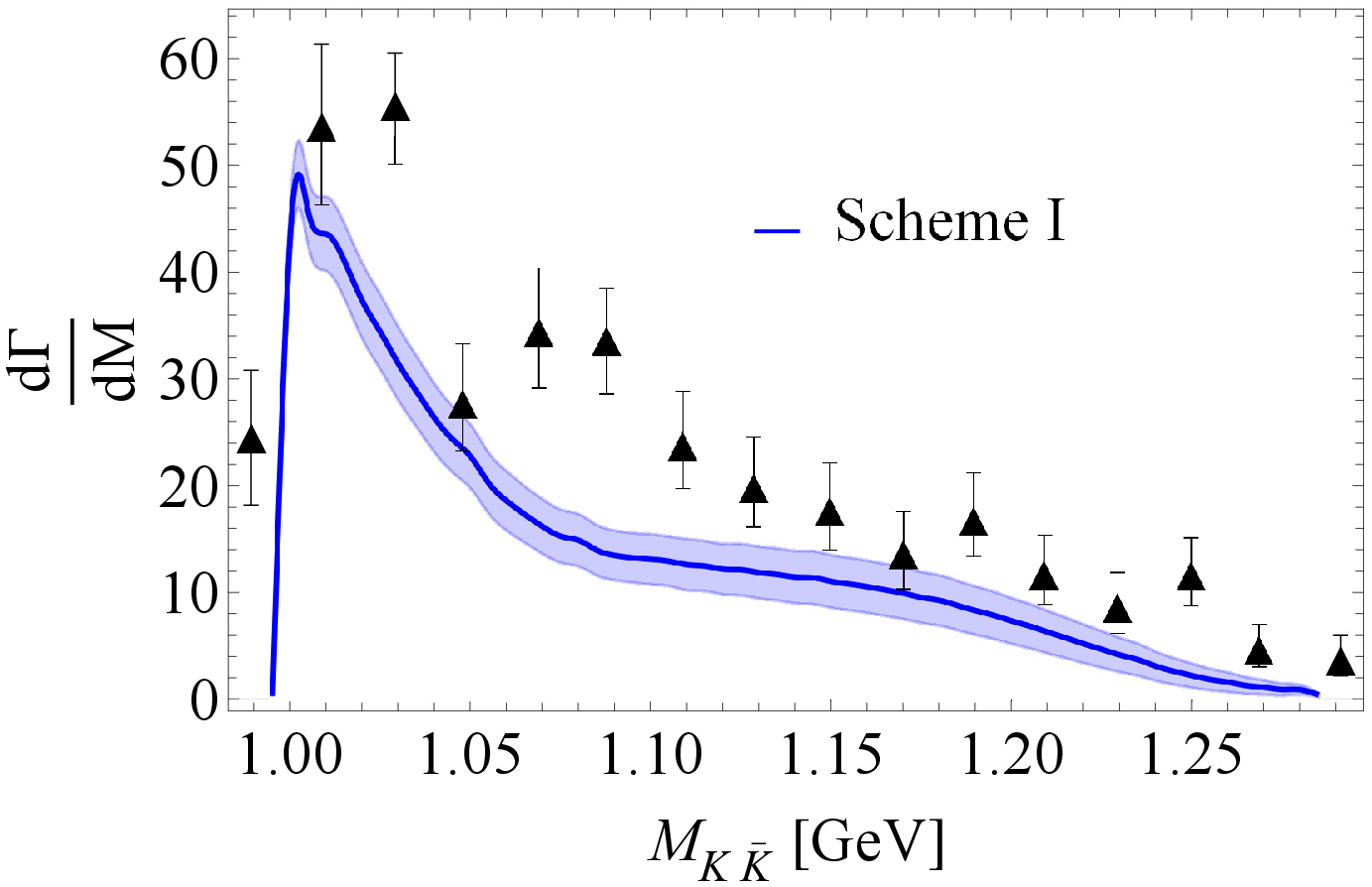}}      \\
  \subfigure[]{\includegraphics[width=3in]{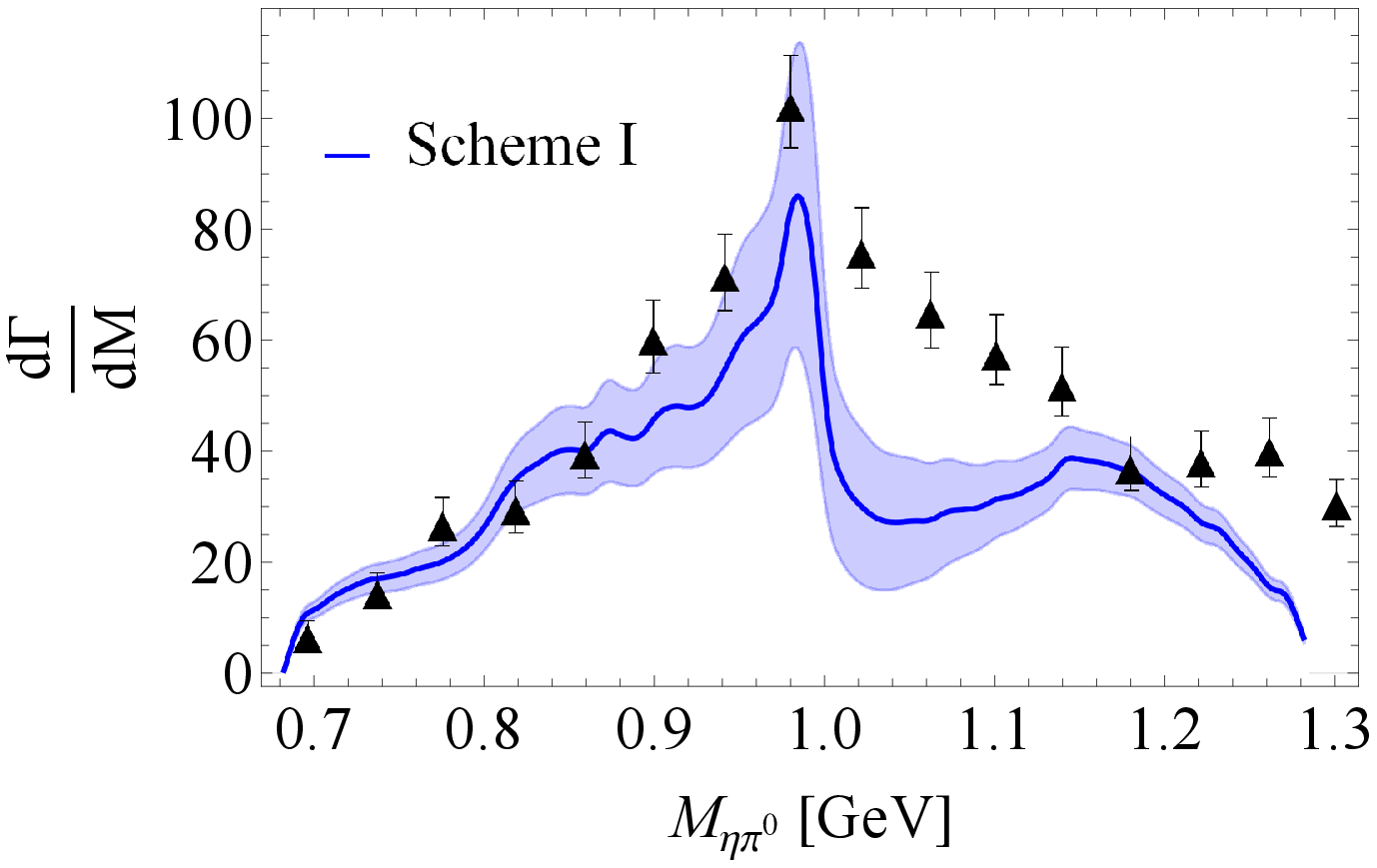}}  \qquad \qquad 
  \subfigure[]{\includegraphics[width=3in]{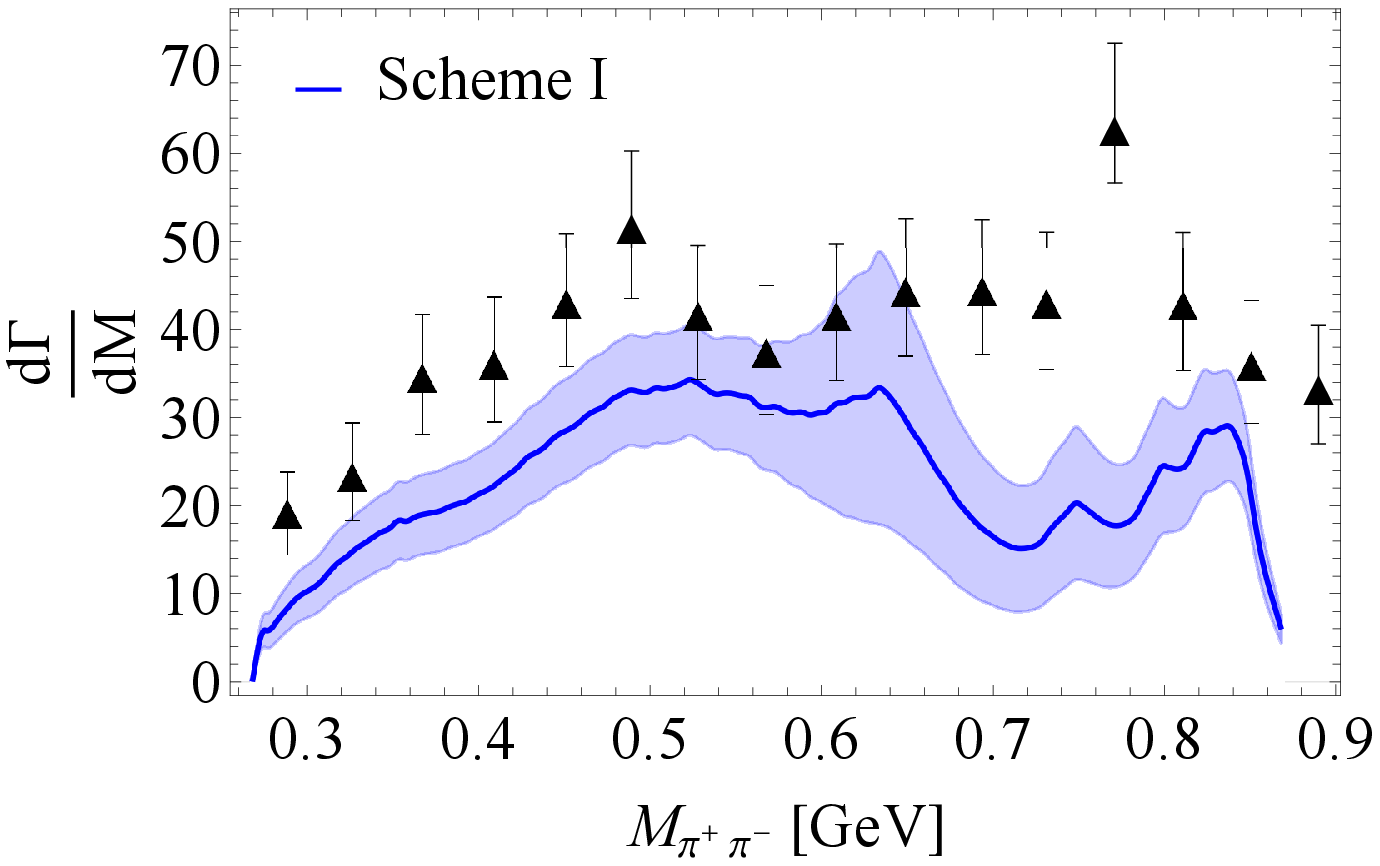}} 

  \caption{Invariant mass spectra in $\eta(1405)$ decay to $K\bar{K}\pi$ or $\eta \pi \pi$. The mass spectra data for $K \pi$ and $K \bar{K}$ denoted by solid triangles are measured in Ref.~\cite{BES:1998bgh}.
  The mass spectra data for $\eta \pi$ and $\pi^+\pi^-$ denoted by solid triangles are measured in Ref.~\cite{BES:1999axp}.}
   \label{Fig:IMS1405-schemeI}
\end{figure}

\begin{figure}
  \centering
  \subfigure[]{\includegraphics[width=3in]{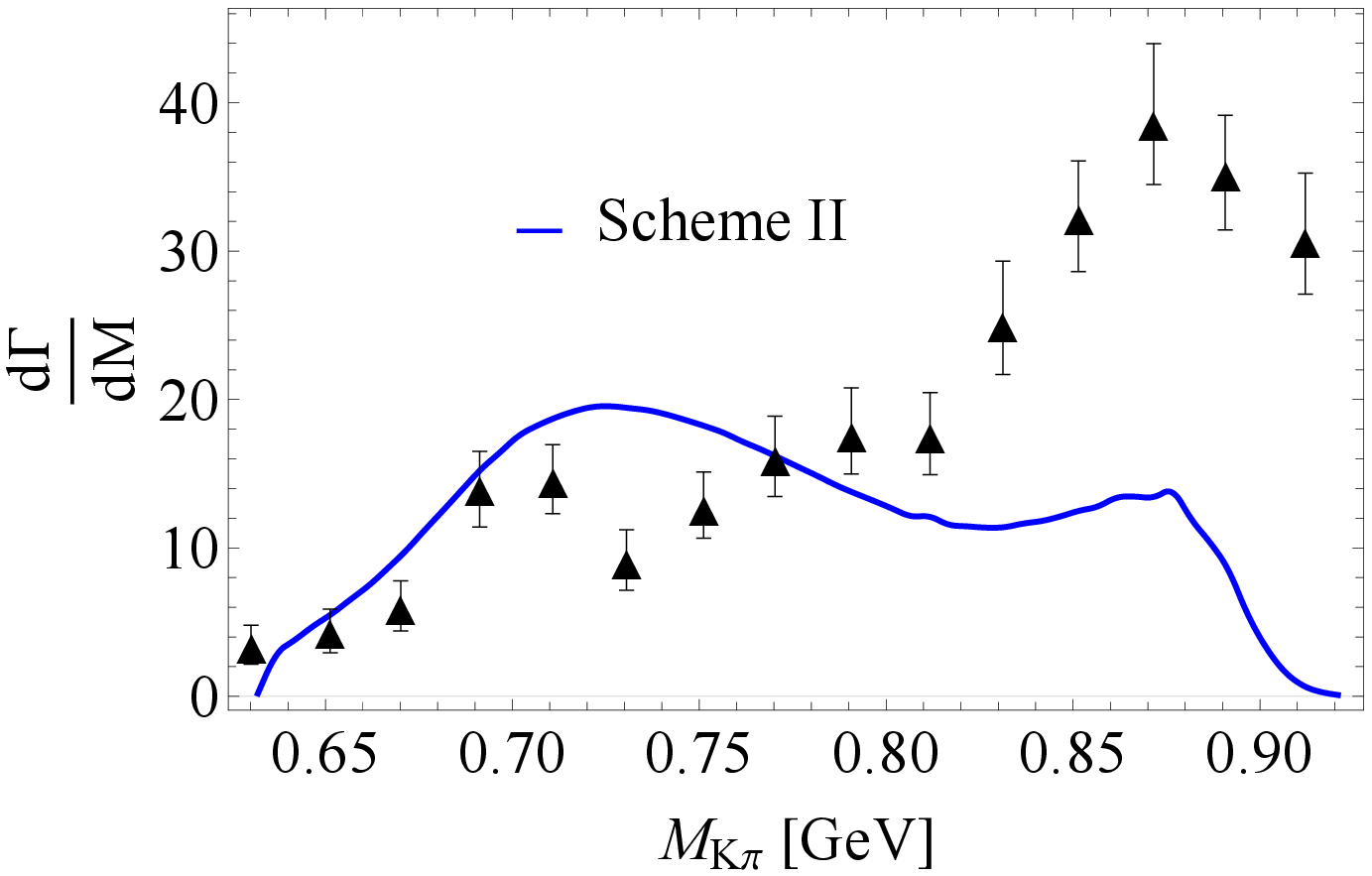}} \qquad \qquad
  \subfigure[]{\includegraphics[width=3in]{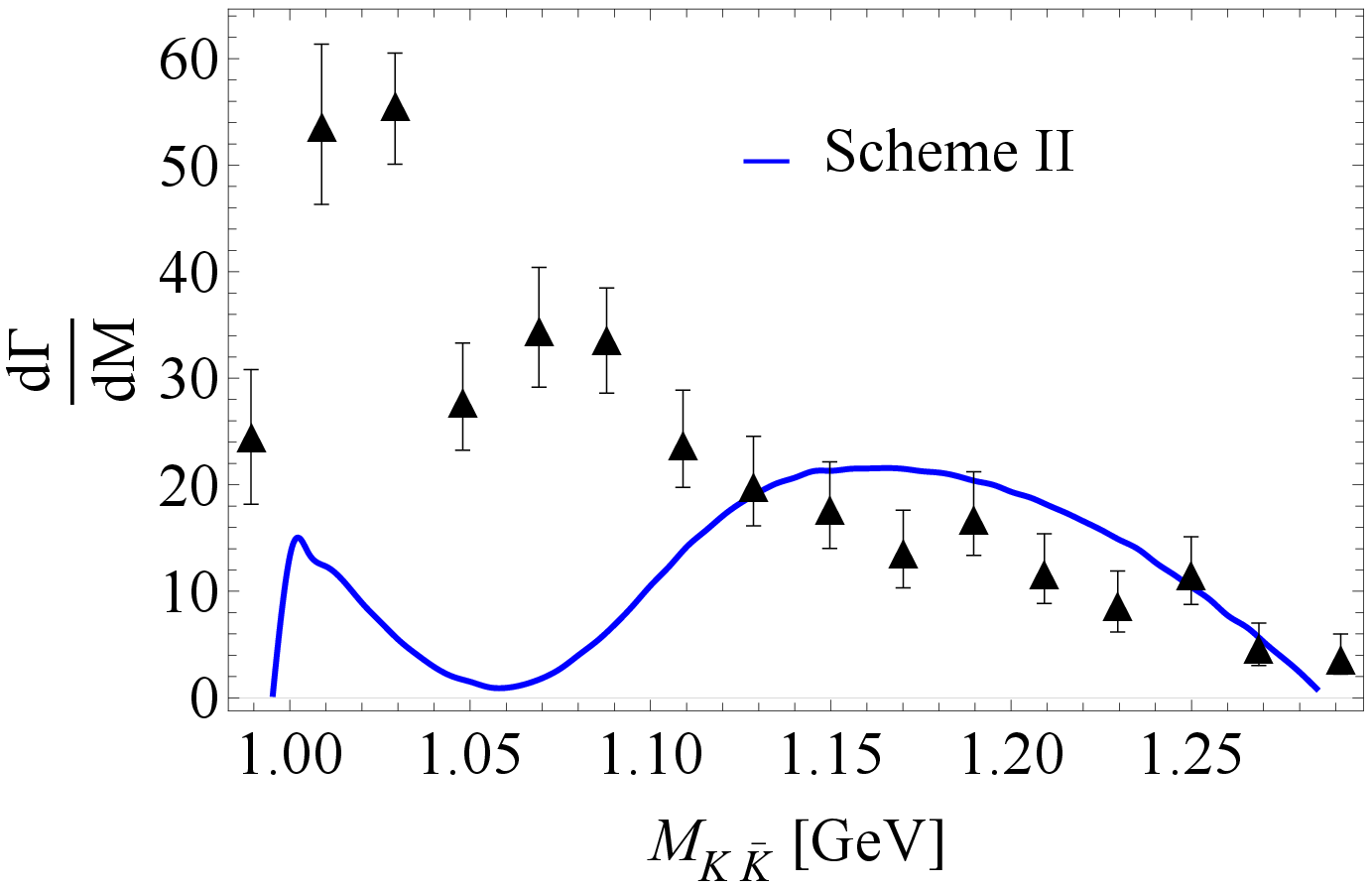}}      \\
  \subfigure[]{\includegraphics[width=3in]{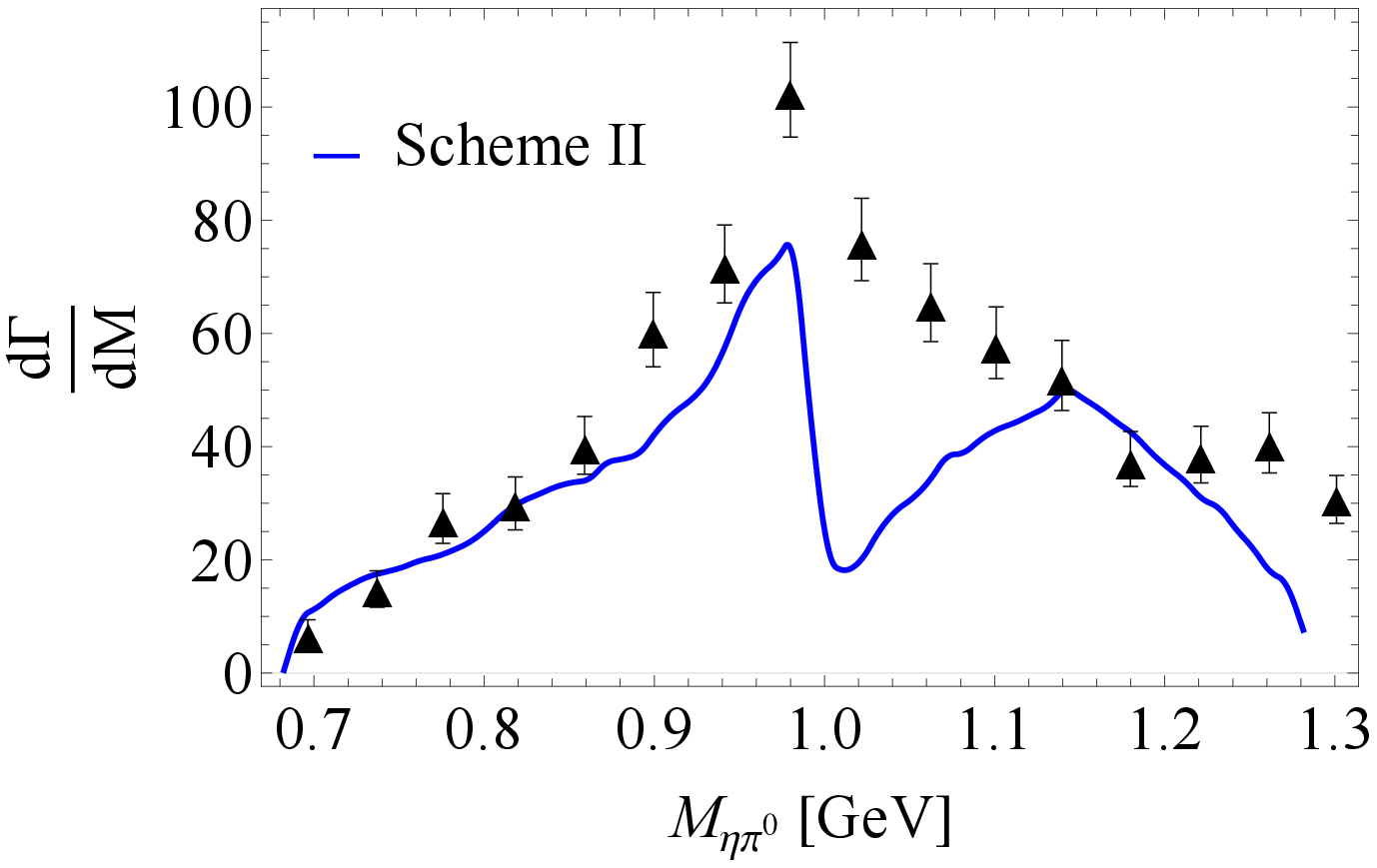}}  \qquad \qquad 
  \subfigure[]{\includegraphics[width=3in]{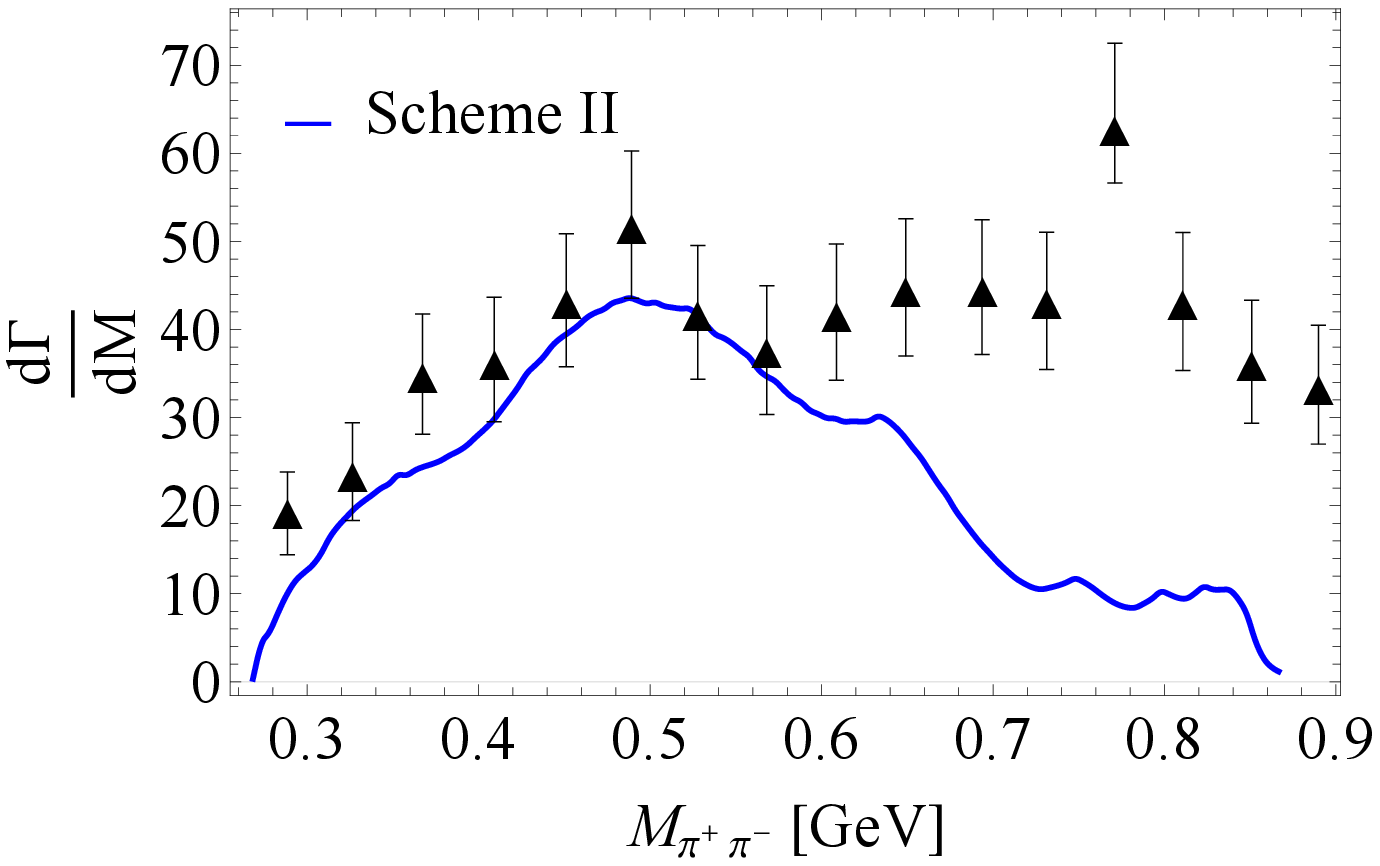}} 

  \caption{Invariant mass spectra of the $\eta(1405)$ decays into $K\bar{K}\pi$ or $\eta \pi \pi$. The experimental data denoted by solid triangles for the $K \pi$ and $K \bar{K}$ invariant mass spectra are from Ref.~\cite{BES:1998bgh}, while the data for the $\eta \pi$ and $\pi^+\pi^-$ invariant mass spectra are from Ref.~\cite{BES:1999axp}.}
   \label{Fig:IMS1405-schemeII}
\end{figure}

The invariant mass spectra can offer strong constraints on the parameters. In Figs.~\ref{Fig:IMS1405-schemeI} and ~\ref{Fig:IMS1405-schemeII} the invariant mass spectra for $K\pi$, $K\bar{K}$, $\eta \pi$ and $\pi \pi$ in the $\eta(1405)$ decays with the fitted parameters in Scheme-I and Scheme-II, respectively, are illustrated.
The solid triangle in Figs.~\ref{Fig:IMS1405-schemeI} and ~\ref{Fig:IMS1405-schemeII} are the experimental data measured by BESII collaboration~\cite{BES:1998bgh,BES:1999axp}. The light blue bands in Fig.~{\ref{Fig:IMS1405-schemeI}} are the errors that comes form the error of the best fitted values in Scheme-I.
While we do not show the error bands in Fig.~\ref{Fig:IMS1405-schemeII} since the error of the parameters in the Scheme-II fit are indeed too large. Thus, we only show the blue lines calculated with the central values of parameters as a comparison with the data.

Comparing the results presented in Figs.~\ref{Fig:IMS1405-schemeI} and ~\ref{Fig:IMS1405-schemeII} we see that the spectra in Scheme-I can better describe the data while the spectra in Scheme-II have large deviations. In particular, the spectra in Fig.~\ref{Fig:IMS1405-schemeII}(a) and (b) cannot match the structure caused by $K^*$. As shown by Tab.~\ref{Tab:fittedparas} the fitted coupling strengths and relative phase angles suggest strong interfering effects arising from different transition mechanisms in the $K\bar{K}\pi$ decay channel. 

For the case of $\eta \pi$, a destructive interference between the tree and triangle loop amplitudes in $\eta(1405)\to a_0 \pi \to \eta \pi \pi$ is favored. Meanwhile, it ensures the $a_0$ contribution to manifest as a peak as shown in Fig.~\ref{Fig:IMS1405-schemeI}(c), instead of a dip as shown in Fig.~\ref{Fig:IMS1405-schemeII}(c). 

\begin{figure}
  \centering
  \subfigure[]{\includegraphics[width=3in]{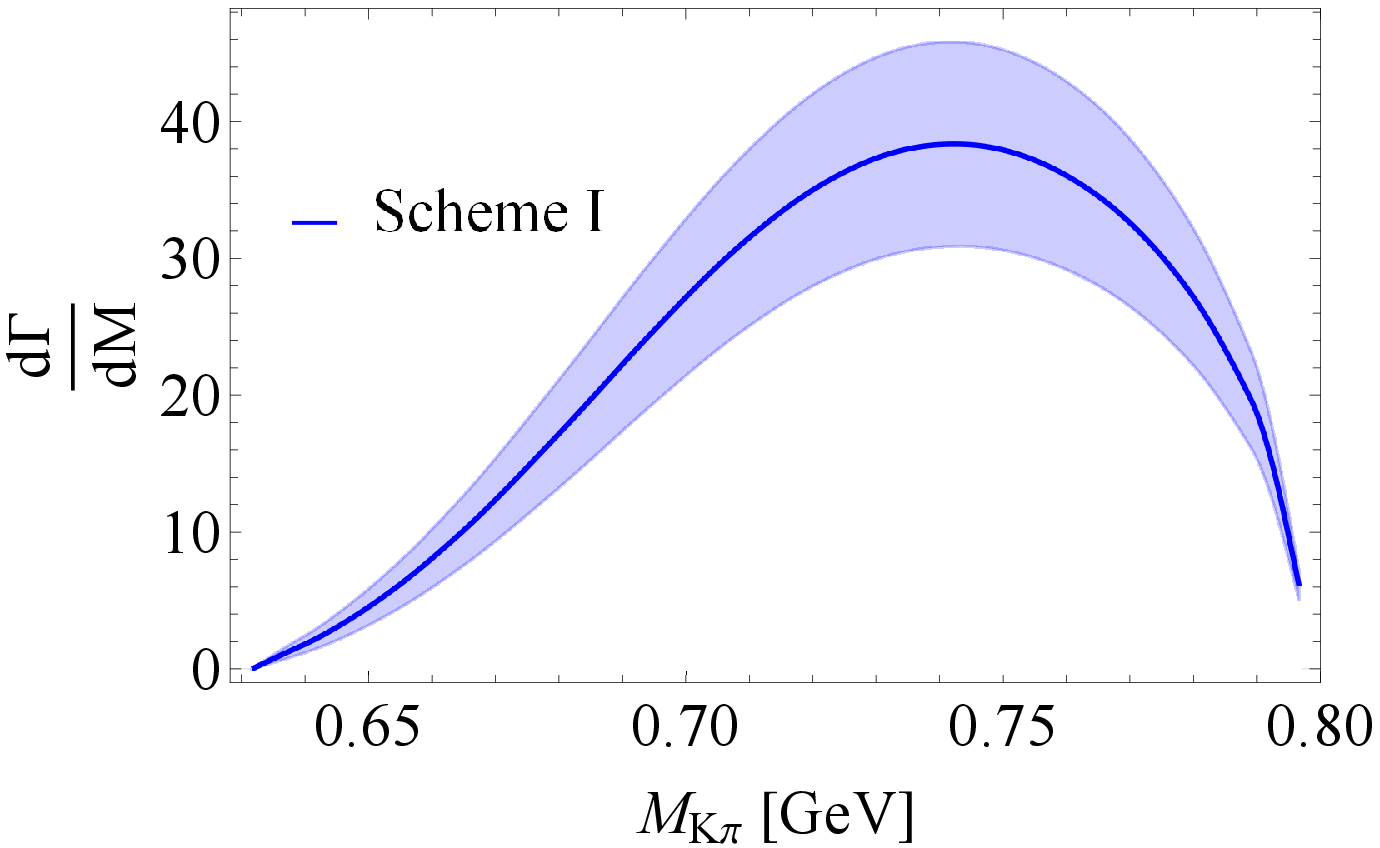}} \qquad \qquad
  \subfigure[]{\includegraphics[width=3in]{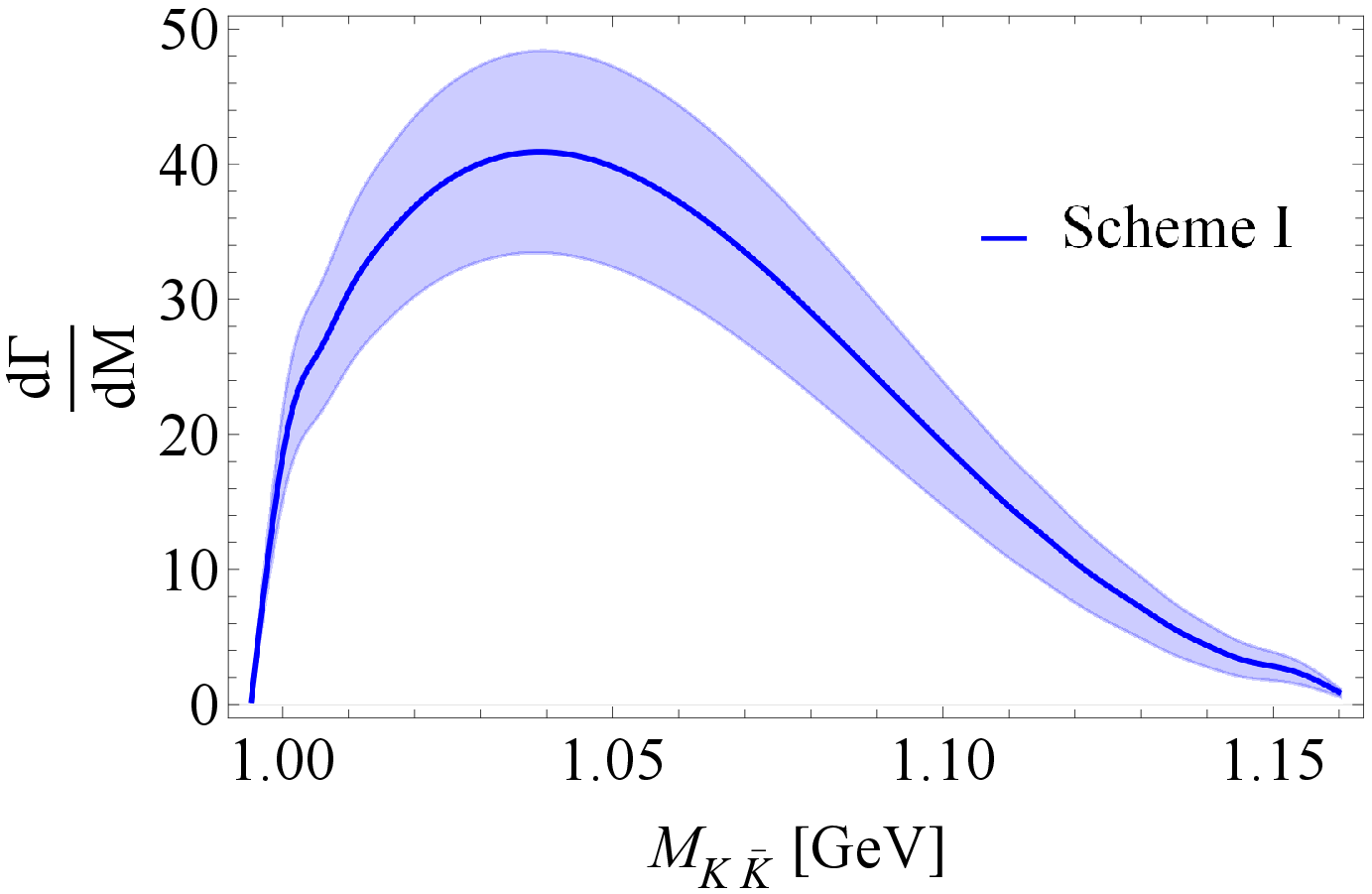}}      \\
  \subfigure[]{\includegraphics[width=3in]{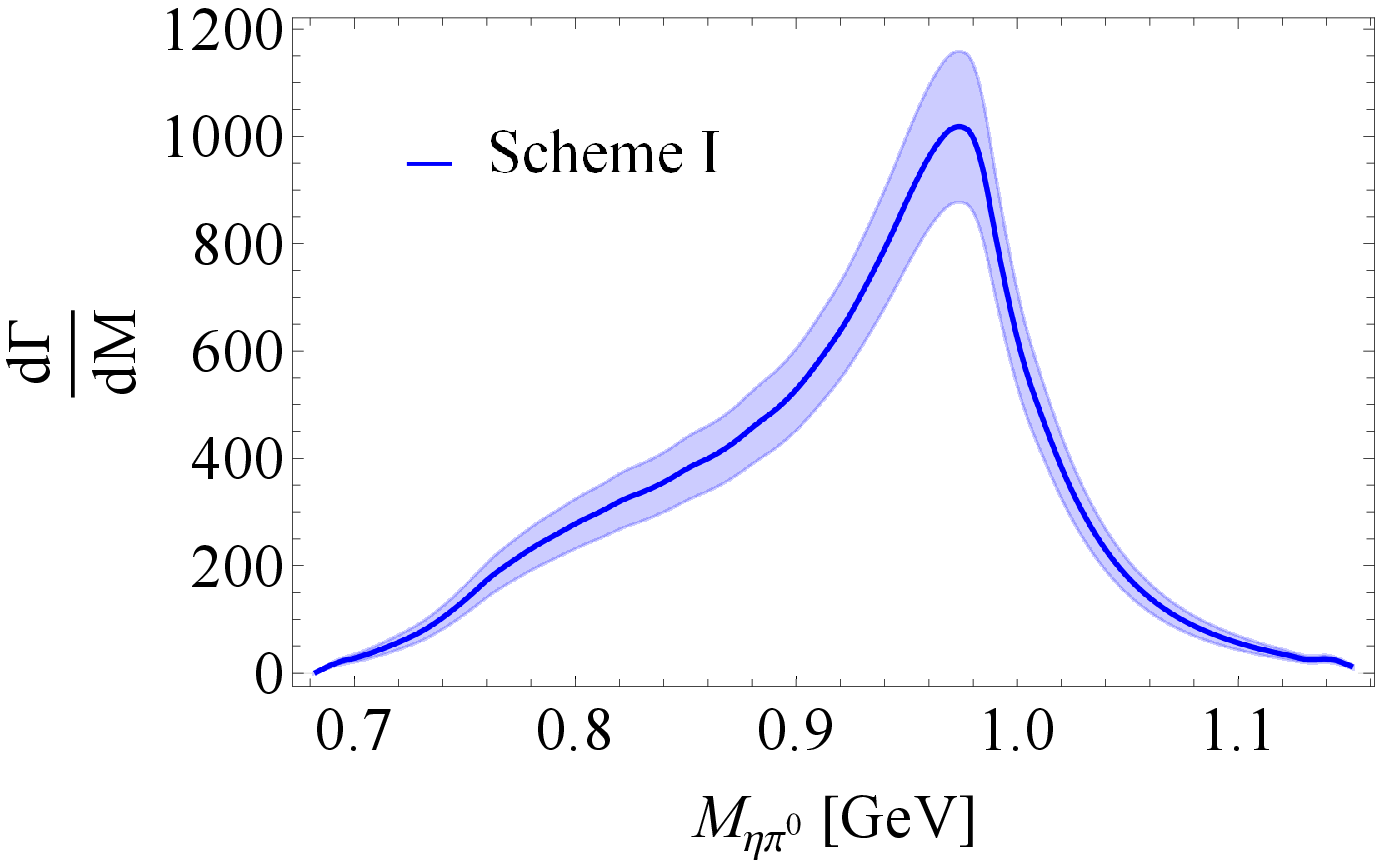}}  \qquad \qquad 
  \subfigure[]{\includegraphics[width=3in]{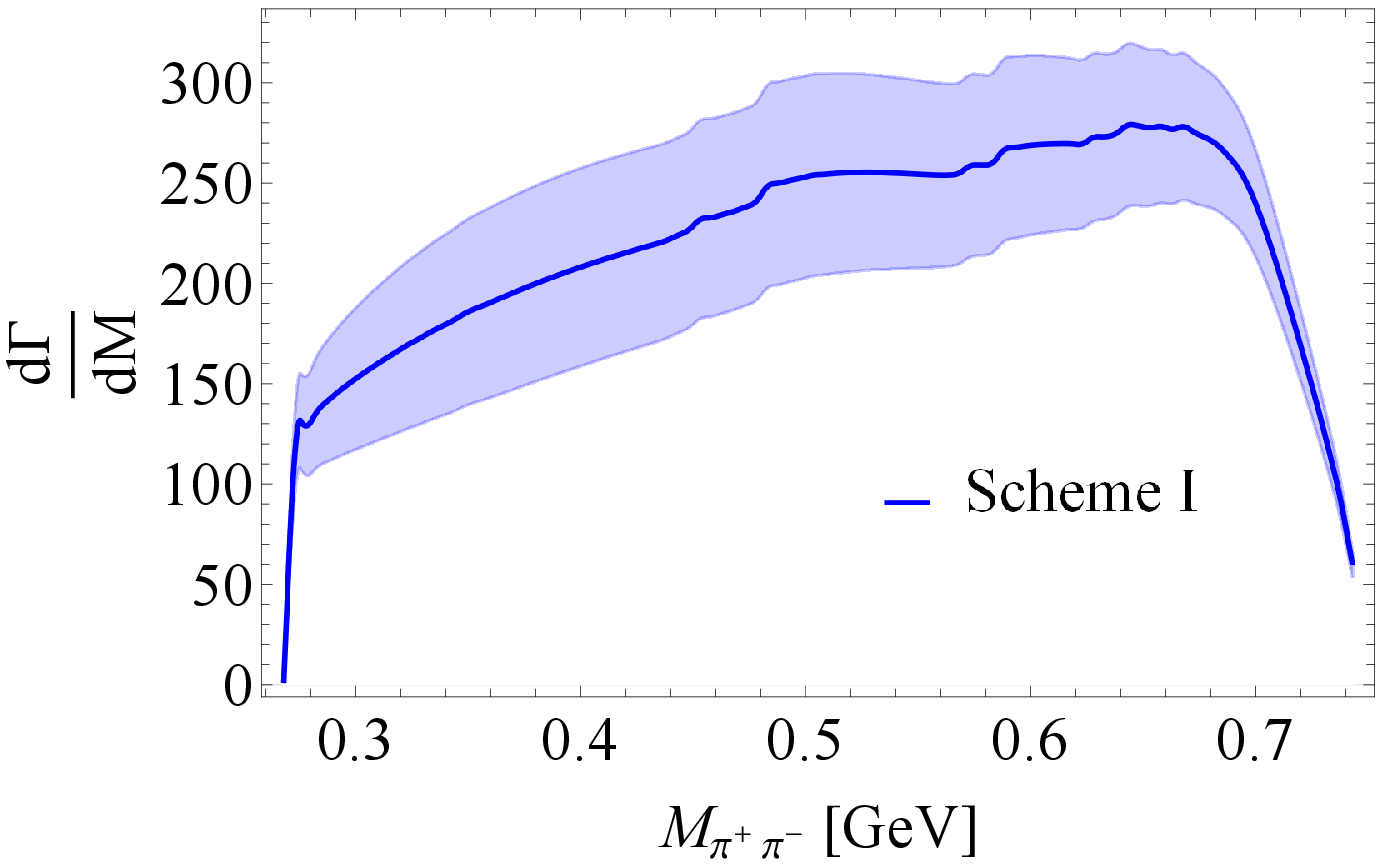}} 

  \caption{Invariant mass spectra of the $\eta(1295)$ decays with the best fitted parameters in Scheme-I. }
   \label{Fig:IMS1295-schemeI}
\end{figure}

In short, the best fitted parameters in Scheme-I are more reasonable for providing an overall description of the available experimental data. As a prediction of our model, the invariant mass spectra of $\eta(1295)$ decaying into $K\bar{K}\pi$ and $\eta \pi \pi$ are illustrated in Fig.~\ref{Fig:IMS1295-schemeI}. 
As shown by Fig.~\ref{Fig:IMS1295-schemeI}(a) the $K^*$ structure  (tree-level contributions from the $K^*\bar{K}+c.c.$ transitions) seems to be submerged by the $S$-wave contributions, i.e., via $\kappa K$ and $a_0(980)\pi$ transitions. We note that a threshold enhancement due to the $a_0(980)$ production is present in the $K\bar{K}$ spectrum in Fig.~\ref{Fig:IMS1295-schemeI}(b). This feature can be quantified by the exclusive calculations of the partial widths from different transition processes later in Tab.~\ref{table:kkpi}.

In the $\eta\pi\pi$ decay channel the dominance of the $S$-wave transition via the $a_0(980)\pi$ is evident as shown by Fig.~\ref{Fig:IMS1295-schemeI}(c).  A broad $\pi\pi$ spectrum in Fig.~\ref{Fig:IMS1295-schemeI}(d) is due to the projection of $a_0(980)\pi\to \eta\pi\pi$ and $\eta\sigma\to \eta\pi\pi$. Also, more quantitative understanding of these features can learned from the exclusive calculations of the partial widths from different transition processes in Tab.~\ref{table:etapipi}.

\subsubsection{Partial widths}

With the parameters determined in Scheme-I, we calculate the exclusive tree-level and loop  contributions to the $K \bar{K}\pi$ and $\eta \pi \pi $ channels, and collect them in Tabs.~\ref{table:kkpi} and ~\ref{table:etapipi}, respectively. To see the sensitivity of the mass of $\eta(1405)$, we calculate two mass values for $\eta(1405)$, i.e., $1.405$ and 1.420 GeV, as a comparison.
The exclusive contributions from different transition processes and the total contributions to $K\bar{K} \pi $ and $\eta \pi \pi $ are collected in Tab.~\ref{Tab:decaywidths}.


\begin{table}
  \centering
  \caption{Exclusive partial widths (in MeV) from the tree and loop-level transitions in $\eta_X \to K\bar{K}\pi$ with the central values of the Scheme-I parameters adopted. The partial widths of the $\eta(1405)$ decays are calculated with two masses $1.405/1.42$ GeV. }
  \begin{tabular}{ll|c|c|c}
  \hline\hline
  \multicolumn{2}{l|}{\multirow{2}{*}{ Partial widths (MeV)}}                                                           & \multicolumn{1}{l|}{\multirow{2}{*}{$\eta(1295)$}} & \multicolumn{2}{c}{$\eta(1405)$} \\ \cline{4-5} 
  \multicolumn{2}{l|}{}                                                                                                 & \multicolumn{1}{l|}{}                              &  $m=1.405$ GeV      & $m=1.42$ GeV           \\ \cline{1-5}
  \multicolumn{1}{l|}{\multirow{3}{*}{Tree level}} & $\eta_X \to K^* \bar{K} \to K \bar{K} \pi $                        &                                    $4.14$&$17.50$&$29.00$        \\
  \multicolumn{1}{l|}{}                            &  $\eta_X \to \kappa \bar{K}  \to K \bar{K} \pi$                    &                                    $0.08$&$3.39$&$3.61$         \\
  \multicolumn{1}{l|}{}                            &  $\eta_X \to  a_0(980) \pi \to  K \bar{K} \pi $                    &                                    $0.27$&$0.36$&$0.38$          \\ \hline\hline
  \multicolumn{1}{l|}{\multirow{2}{*}{Loop level}} & $\eta_X \to K^* \bar{K} \to  K \bar{\kappa} \to  K (\bar{K}\pi)$   &                                    $15.41$&$9.64$&$11.12$       \\
  \multicolumn{1}{l|}{}                            &  $\eta_X \to K^* \bar{K} \to a_0(980) \pi  \to (K \bar{K} ) \pi $  &                                    $7.31$&$1.91$&$2.05$          \\ \hline \hline
 \end{tabular}
 \label{table:kkpi}
  \end{table}


\begin{table}
  \centering
  \caption{Exclusive partial widths (in MeV) from the tree and loop-level transitions  in $\eta_X \to \eta\pi\pi$ with the central values of the Scheme-I parameters adopted. The partial widths of the $\eta(1405)$ decays are calculated with two masses $1.405/1.42$ GeV. }
  \begin{tabular}{ll|c|c|c}
  \hline\hline
  \multicolumn{2}{l|}{\multirow{2}{*}{ Channels (MeV)}}                                                                  & \multicolumn{1}{l|}{\multirow{2}{*}{$\eta(1295)$}} & \multicolumn{2}{c}{$\eta(1405)$} \\ \cline{4-5} 
  \multicolumn{2}{l|}{}                                                                                                  & \multicolumn{1}{l|}{}                              &  $m=1.405$ GeV      & $m=1.42$ GeV           \\ \cline{1-5}
  \multicolumn{1}{l|}{\multirow{3}{*}{Tree level}} & $\eta_X \to \sigma \eta \to \eta \pi \pi  $                         &                                           $0.62$&$0.55$&$0.56$       \\
  \multicolumn{1}{l|}{}                            &  $\eta_X \to f_0(980)\eta \to \eta \pi \pi $                        &                                           $0.03$&$0.06$&$0.06$     \\
  \multicolumn{1}{l|}{}                            &  $\eta_X \to a_0(980) \pi \to \eta \pi \pi $                        &                                           $3.24$&$3.03$&$3.07$       \\ \hline\hline
  \multicolumn{1}{l|}{\multirow{2}{*}{Loop level}} &  $\eta_X \to K^*\bar{K} \to \sigma \eta \to  \eta \pi \pi $         &                                           $0.68$&$0.15$&$0.21$       \\
  \multicolumn{1}{l|}{}                            &  $\eta_X \to K^* \bar{K} \to \eta f_0(980)  \to \eta \pi \pi $      &                                           $0.83$&$0.47$&$0.63$  \\ 
  \multicolumn{1}{l|}{}                            & $\eta_X \to K^* \bar{K} \to a_0(980) \pi  \to \eta \pi \pi $        &                                           $28.00$&$4.71$&$5.60$    \\ \hline \hline
\end{tabular}
 \label{table:etapipi}
  \end{table}


Compared to the $\eta(1405)$ decay, the width of $\eta(1295)$ decaying to $K\bar{K}\pi$ will be suppressed by the limited phase space. At the tree level, the mass of $\eta(1295)$ is below the thresholds of $K^*\bar{K}$ and $\kappa \bar{K}$, which means that its couplings to these two-body decay channels will be reflected by the three-body $K\bar{K}\pi$ decays. Note that $\eta(1295)$ has a large coupling to $K^* \bar{K}$. It implies that the intermediate $K^* \bar{K}$ may produce crucial interference effects at high mass regions. The other important consequence is that due to the TS mechanism the $K^* \bar{K}$ rescattering via the triangle loops may contribute significantly to the productions of $\kappa\bar{K}$ and $a_0(980)\pi$ which will then decay into $K\bar{K}\pi$. For the $\eta(1295)$ decays, although it does not satisfies the TS condition, the large $\eta(1295)$ coupling to $K^*\bar{K}$ enhances the triangle loop contributions. For the $\eta(1405)$ decays, although its coupling to $K^*\bar{K}$ is much smaller than that of $\eta(1295)$, the satisfaction of the kinematic condition for the TS mechanism enhances its contributions to the $\kappa\bar{K}$ and $a_0(980)\pi$ productions. Eventually, we find that the loop contributions are even larger than the tree-level ones, where the bare $\eta_X$ coupling to $\kappa\bar{K}$ is not a free parameter but related to the bare $a_0(980)\pi$ via the flavor SU(3) relation as given in Tab.~\ref{tab:xspcouplings}. We stress that this assumption is questionable and should be examined by future experiment. At this moment, there is no better way to constrain it.

For the $\eta \pi \pi $ channel, as shown in Tab.~\ref{table:etapipi}, the most significant tree-level contribution is from the $a_0(980) \pi$ channel. Note that the $\sigma \eta $ and $f_0 \eta $ channels do not satisfy the TS mechanism while the $a_0(980) \pi$ does. The TS mechanism has strongly enhanced the $a_0(980) \pi$ contributions to the $\eta\pi\pi$ channel. This result seems to be in contradiction with some of the early measurements~\cite{E852:2000rhq,GAMS:1997pxg,Anisovich:2001jb,BES:1999axp} where large contributions from the $\sigma \eta$ channel were reported. Again, this may be due to the treatment of the $\eta_X SP$ couplings in the SU(3) flavor symmetry.  

In Tab.~\ref{Tab:decaywidths} we combine the tree and loop amplitude for the intermediate two-body decays of $\eta(1295)$ and $\eta(1405)$ into either $K\bar{K}\pi$ and $\eta\pi\pi$. It shows that both $\eta(1295)$ and $\eta(1405)$ can have significant partial decay widths into $K\bar{K}\pi$. This feature suggests that a coherent analysis of the $K\bar{K}\pi$ channel including both $\eta(1295)$ and $\eta(1405)$ should be necessary since their interferences may produce some measurable effects. We will show in an forthcoming work that it is a key for understanding the recent BESIII measurement of the $K\bar{K}\pi$ spectrum in the $J/\psi$ radiative decays~\cite{BESIII:2022chl}. 

From Tab.~\ref{Tab:decaywidths} one can see that the four-body decays do not have significant contributions to the total widths of $\eta(1295)$ due to the subthreshold $P$-wave couplings for $\eta(1295)\to VV$. In contrast, the $4\pi$ decay width of $\eta(1405)$ turns out to be sizeable and may be searched in experiment.

\begin{table}
  \centering
  \caption{The exclusive partial widths of $\eta_X$ decays into $K\bar{K}\pi$, $\eta \pi \pi$, $K\bar{K}\pi\pi$ and $4\pi$. For the $K\bar{K}\pi$ and $\eta \pi \pi$ decay channels, the tree and loop amplitudes are combined to the intermediate two-body decay partial waves. Partial widths for the $\eta(1405)$ case are calculated for two masses $1.405/1.42$ GeV. For the quantity which is too small we do not show the error.}
 \begin{tabular}{l|c |cc}
      \hline \hline
        \multirow{2}*{ Channels (MeV) }                                  &   \multirow{2}*{$\eta(1295)$ }  &   \multicolumn{2}{c}{ $\eta(1405)$}       \\ \cline{3-4}
                                                                         &                                 &   m=$1.405$ GeV   &  m=$1.42$ GeV        \\ \hline
       $\eta_X \to K^*\bar{K}  \to  K\bar{K}\pi $                        &   $4.14\pm0.03$&$17.51\pm0.21$&$29.00\pm0.31$ \\          
                                                                        
      $\eta_X \to \kappa \bar{K} \to K \bar{K} \pi $                     &   $17.00\pm1.51$&$21.51\pm1.51$&$24.20\pm1.72$             \\ 
                                                                        
       $\eta_X \to a_0(980) \pi  \to K\bar{K} \pi $                      &   $9.86\pm0.68$&$1.02\pm0.12$&$1.15\pm0.13$         \\ \hline

      $\eta_X  \to K\bar{K} \pi $                                        &   $38.90\pm2.61$&$30.40\pm1.81$&$40.61\pm2.12$             \\ \hline \hline

      $\eta_X \to \sigma \eta \to \eta \pi \pi $                         &    $0.20\pm0.04$&$0.94\pm0.06$&$1.02\pm0.11$             \\

      $\eta_X \to f_0(980) \eta \to \eta \pi \pi$                        &    $1.12\pm0.11$&$0.76\pm0.06$&$0.97\pm0.06$             \\

      $\eta_X \to a_0(980) \pi \to \eta \pi \pi $                        &    $46.10\pm2.81$&$3.33\pm0.32$&$4.05\pm0.43$              \\ \hline

        $\eta_X  \to \eta \pi \pi $                                      &   $51.81\pm3.12$&$5.09\pm0.41$&$6.75\pm0.58$               \\ \hline \hline
        $\eta_X \to K \bar{K} \pi \pi $                                  &   $1.60 \times 10^{-11}$ &  $5.00 \times 10^{-5}$&  $1.10 \times 10^{-4}$                    \\
        $\eta_X \to  4 \pi  $                                            &   $0.98 \pm 0.31$& $4.12\pm1.31$ & $5.12\pm1.61$                 \\ \hline \hline     
  \end{tabular}
   \label{Tab:decaywidths}
\end{table}

In brief, the width of $\eta(1405)$ is dominated by the $K\bar{K}\pi$ channel, and $\eta(1295)$ has both the $K\bar{K}\pi$ and $\eta \pi \pi $ channels to contribute dominantly to the total width via the triangle loop transitions. It should be mentioned that sensitivities of the results to the cut-off parameter $\Lambda$ for the triangle loops do exist. It implies that a combined analysis including the TS mechanism is necessary for both the $K\bar{K}\pi$ and $\eta\pi\pi$ channel. We should also give sufficient cautions on the treatment of the $\eta_XSP$ couplings where the SU(3) flavor symmetry is applied. Taking into account the different properties of these scalars below 1 GeV, these couplings may strongly deviate from the SU(3) relation. Future high-statistics data are required for a better constraint on these quantities.

\section{Summary}\label{sec:4}

In summary, we have made a systematic study of the total widths of $\eta(1295)$ and $\eta(1405)$ as the first radial excitation states of $\eta$ and $\eta'$. By fitting the available B.R. fraction data for the $\eta(1405/1475)$ decays, we can determine the parameters and make a reasonable description of the invariant mass spectra in both the $\eta(1295)$ and $\eta(1405)$ decays into $K\bar{K}\pi$ and $\eta\pi\pi$. Our study suggests that these two channels are the dominant ones which involve several intermediate two-body transitions, such as $\kappa \bar{K}$, $a_0(980) \pi $, $\sigma \eta $ and $f_0(980) \eta$. In particular, our study shows that the TS mechanism plays a crucial role in the productions of $\kappa \bar{K}$ and $a_0(980) \pi$ in the $K\bar{K}\pi$ channel, and in the production of $a_0(980) \pi$ in the $\eta\pi\pi$ channel. This is in agreement with our previous findings. 

We have also studied the four-body decays into $K\bar{K}\pi\pi$ and $4\pi$ for $\eta(1295)$ and $\eta(1405)$ and find they are rather small. It confirms that the three-body decays into $K\bar{K}\pi$ and $\eta\pi\pi$ have nearly saturated the total widths of these two states. Note that the interferences due to different intermediate two-body transitions are important. It suggests that a combined partial wave analysis from threshold up to 1.5 GeV is necessary. The relevant analysis will be reported in a forthcoming work.

\begin{acknowledgments}
This work is supported, in part, by the National Natural Science Foundation of China (Grant No. 12235018),  DFG and NSFC funds to the Sino-German CRC 110 ``Symmetries and the Emergence of Structure in QCD'' (NSFC Grant No. 12070131001, DFG Project-ID 196253076), National Key Basic Research Program of China under Contract No. 2020YFA0406300, and Strategic Priority Research Program of Chinese Academy of Sciences (Grant No. XDB34030302).
\end{acknowledgments}

\begin{appendix}
\section*{Appendix A: The SU(3) relations for the scalar meson couplings to two pseudoscalar mesons}
As an approximation, we adopt the following SU(3) relations in the determination of the scalar meson couplings to a pair of pseudoscalar mesons:
\begin{itemize}
    \item  $\sigma-\pi \pi $
\begin{eqnarray}
    g_{\sigma \pi^0 \pi^0}= g_{\sigma \pi^- \pi^+}= g_{\sigma \pi^+ \pi^-}=\sqrt{2} g_{SPP} \ , 
  \end{eqnarray}
    \item $a_0 - \eta \pi $
\begin{eqnarray}
    g_{a_0 \eta \pi^0}=g_{a_0^+ \eta \pi^+}=g_{a_0^- \eta \pi^-} =\sqrt{2} \cos \alpha_P g_{SPP} \ ,
\end{eqnarray}

    \item $a_0-K \bar{K}$
\begin{eqnarray}
     g_{a_0 K^+ K^-} =\frac{g_{SPP}}{\sqrt{2}}=-g_{a_0 K^0 \bar{K}^0}= \frac{g_{a_0^+ K^+ \bar{K}^0}}{\sqrt{2}}=\frac{g_{a_0^- K^- K^0}}{\sqrt{2}} \ ,
\end{eqnarray}
  \item $f_0- K \bar{K}$
\begin{eqnarray}
    g_{f_0 K^+ K^-}=g_{f_0 K^0 \bar{K}^0}=g_{SPP} \ .
\end{eqnarray}
\end{itemize}

\end{appendix}

\bibliography{bibfile}

\begin{thebibliography}{10}

\bibitem{Baillon:1967zz}
P.~H. Baillon et~al.
\newblock {Further Study of the e-Meson in Antiproton Proton Annihilation at
  Rest}.
\newblock {\em Nuovo Cim. A}, 50:393--421, 1967.

\bibitem{MARK-III:1990wgk}
Z.~Bai et~al.
\newblock {Partial wave analysis of J / psi ---\ensuremath{>} gamma K0(s) K+-
  pi-+}.
\newblock {\em Phys. Rev. Lett.}, 65:2507--2510, 1990.

\bibitem{DM2:1990cwz}
J.~E. Augustin et~al.
\newblock {Partial wave analysis of DM2 data in the eta (1430) energy range}.
\newblock {\em Phys. Rev. D}, 46:1951--1958, 1992.

\bibitem{OBELIX:2002eai}
F~Nichitiu et~al.
\newblock {Study of the K+ K- pi+ pi- pi0 final state in anti-proton
  annihilation at rest in gaseous hydrogen at NTP with the OBELIX
  spectrometer}.
\newblock {\em Phys. Lett. B}, 545:261--271, 2002.

\bibitem{Note1}
It should be pointed out that the fitted resonance parameters for $\eta (1405)$
  and $\eta (1475)$ by MARKIII~\cite {MARK-III:1990wgk}, DM-2~\cite
  {DM2:1990cwz} and Obelix are not consistent~\cite {OBELIX:2002eai}.

\bibitem{Donoghue:1980hw}
John~F. Donoghue, K.~Johnson, and Bing~An Li.
\newblock {Low Mass Glueballs in the Meson Spectrum}.
\newblock {\em Phys. Lett. B}, 99:416--420, 1981.

\bibitem{Close:1980rv}
F.~E. Close and S.~Monaghan.
\newblock {A New Approach to Interactions in the {MIT} Bag}.
\newblock {\em Phys. Rev. D}, 23:2098, 1981.

\bibitem{Close:1987er}
F.~E. Close.
\newblock {Gluonic Hadrons}.
\newblock {\em Rept. Prog. Phys.}, 51:833, 1988.

\bibitem{Amsler:2004ps}
Claude Amsler and N.~A. Tornqvist.
\newblock {Mesons beyond the naive quark model}.
\newblock {\em Phys. Rept.}, 389:61--117, 2004.

\bibitem{Masoni:2006rz}
A.~Masoni, C.~Cicalo, and G.~L. Usai.
\newblock {The case of the pseudoscalar glueball}.
\newblock {\em J. Phys. G}, 32:R293--R335, 2006.

\bibitem{Rosenzweig:1981cu}
C.~Rosenzweig, A.~Salomone, and J.~Schechter.
\newblock {A Pseudoscalar Glueball, the Axial Anomaly and the Mixing Problem
  for Pseudoscalar Mesons}.
\newblock {\em Phys. Rev. D}, 24:2545--2548, 1981.

\bibitem{Cheng:2008ss}
Hai-Yang Cheng, Hsiang-nan Li, and Keh-Fei Liu.
\newblock {Pseudoscalar glueball mass from eta - eta-prime - G mixing}.
\newblock {\em Phys. Rev. D}, 79:014024, 2009.

\bibitem{Close:1996yc}
Frank~E. Close, Glennys~R. Farrar, and Zhen-ping Li.
\newblock {Determining the gluonic content of isoscalar mesons}.
\newblock {\em Phys. Rev. D}, 55:5749--5766, 1997.

\bibitem{Li:2007ky}
Gang Li, Qiang Zhao, and Chao-Hsi Chang.
\newblock {Decays of J/ psi and psi-prime into vector and pseudoscalar meson
  and the pseudoscalar glueball-q anti-q mixing}.
\newblock {\em J. Phys. G}, 35:055002, 2008.

\bibitem{Gutsche:2009jh}
Thomas Gutsche, Valery~E. Lyubovitskij, and Malte~C. Tichy.
\newblock {eta(1405) in a chiral approach based on mixing of the pseudoscalar
  glueball with the first radial excitations of eta and eta-prime}.
\newblock {\em Phys. Rev. D}, 80:014014, 2009.

\bibitem{Li:2009rk}
Bing~An Li.
\newblock {Chiral field theory of 0-+ glueball}.
\newblock {\em Phys. Rev. D}, 81:114002, 2010.

\bibitem{Tsai:2011dp}
Yu-Dai Tsai, Hsiang-nan Li, and Qiang Zhao.
\newblock {$\eta_c$ mixing effects on charmonium and $B$ meson decays}.
\newblock {\em Phys. Rev. D}, 85:034002, 2012.

\bibitem{Eshraim:2012jv}
Walaa~I. Eshraim, Stanislaus Janowski, Francesco Giacosa, and Dirk~H. Rischke.
\newblock {Decay of the pseudoscalar glueball into scalar and pseudoscalar
  mesons}.
\newblock {\em Phys. Rev. D}, 87(5):054036, 2013.

\bibitem{Chen:2005mg}
Y.~Chen et~al.
\newblock {Glueball spectrum and matrix elements on anisotropic lattices}.
\newblock {\em Phys. Rev. D}, 73:014516, 2006.

\bibitem{Bali:1993fb}
G.~S. Bali, K.~Schilling, A.~Hulsebos, A.~C. Irving, Christopher Michael, and
  P.~W. Stephenson.
\newblock {A Comprehensive lattice study of SU(3) glueballs}.
\newblock {\em Phys. Lett. B}, 309:378--384, 1993.

\bibitem{Morningstar:1999rf}
Colin~J. Morningstar and Mike~J. Peardon.
\newblock {The Glueball spectrum from an anisotropic lattice study}.
\newblock {\em Phys. Rev. D}, 60:034509, 1999.

\bibitem{Chowdhury:2014mra}
Abhishek Chowdhury, A.~Harindranath, and Jyotirmoy Maiti.
\newblock {Correlation and localization properties of topological charge
  density and the pseudoscalar glueball mass in SU(3) lattice Yang-Mills
  theory}.
\newblock {\em Phys. Rev. D}, 91(7):074507, 2015.

\bibitem{BESIII:2013cbb}
M.~Ablikim et~al.
\newblock {Study of $\psi(3686) \rightarrow \omega K \bar{K} \pi$ decays}.
\newblock {\em Phys. Rev. D}, 87:092006, 2013.

\bibitem{BESIII:2011nqb}
M.~Ablikim et~al.
\newblock {$\eta\pi^+\pi^-$ Resonant Structure around 1.8 GeV/$c^2$ and
  $\eta(1405)$ in $J/psi\to \omega\eta\pi^+\pi^-$}.
\newblock {\em Phys. Rev. Lett.}, 107:182001, 2011.

\bibitem{BESIII:2010gmv}
M.~Ablikim et~al.
\newblock {Confirmation of the $X(1835)$ and observation of the resonances
  $X(2120)$ and $X(2370)$ in $J/\psi\to \gamma \pi^+\pi^-\eta^\prime$}.
\newblock {\em Phys. Rev. Lett.}, 106:072002, 2011.

\bibitem{BESIII:2019yzg}
Medina Ablikim et~al.
\newblock {Measurement of branching fractions of $\psi(3686)\to
  \phi\eta^\prime, \phi f_1(1285)$ and $\phi \eta(1405)$}.
\newblock {\em Phys. Rev. D}, 100(9):092003, 2019.

\bibitem{BESIII:2012aa}
M.~Ablikim et~al.
\newblock {First observation of $\eta(1405)$ decays into $f_{0}(980)\pi^0$}.
\newblock {\em Phys. Rev. Lett.}, 108:182001, 2012.

\bibitem{Mathieu:2009sg}
Vincent Mathieu and Vicente Vento.
\newblock {Pseudoscalar glueball and eta - eta-prime mixing}.
\newblock {\em Phys. Rev. D}, 81:034004, 2010.

\bibitem{Qin:2017qes}
Wen Qin, Qiang Zhao, and Xian-Hui Zhong.
\newblock {Revisiting the pseudoscalar meson and glueball mixing and key issues
  in the search for a pseudoscalar glueball state}.
\newblock {\em Phys. Rev. D}, 97(9):096002, 2018.

\bibitem{Li:2021gsx}
Hsiang-nan Li.
\newblock {Dispersive analysis of glueball masses}.
\newblock {\em Phys. Rev. D}, 104(11):114017, 2021.

\bibitem{Landau:1959fi}
L.~D. Landau.
\newblock {On analytic properties of vertex parts in quantum field theory}.
\newblock {\em Nucl. Phys.}, 13(1):181--192, 1959.

\bibitem{Cutkosky:1960sp}
R.~E. Cutkosky.
\newblock {Singularities and discontinuities of Feynman amplitudes}.
\newblock {\em J. Math. Phys.}, 1:429--433, 1960.

\bibitem{Wu:2011yx}
Jia-Jun Wu, Xiao-Hai Liu, Qiang Zhao, and Bing-Song Zou.
\newblock {The Puzzle of anomalously large isospin violations in
  $\eta(1405/1475)\to 3\pi$}.
\newblock {\em Phys. Rev. Lett.}, 108:081803, 2012.

\bibitem{Note2}
In this work we refer the single state as either $\eta (1405/1475)$ or just
  $\eta (1405)$ if it does not bring confusions.

\bibitem{Wu:2012pg}
Xiao-Gang Wu, Jia-Jun Wu, Qiang Zhao, and Bing-Song Zou.
\newblock {Understanding the property of $\eta(1405/1475)$ in the $J/\psi$
  radiative decay}.
\newblock {\em Phys. Rev. D}, 87(1):014023, 2013.

\bibitem{Aceti:2012dj}
F.~Aceti, W.~H. Liang, E.~Oset, J.~J. Wu, and B.~S. Zou.
\newblock {Isospin breaking and $f_0(980)$-$a_0(980)$ mixing in the $\eta(1405)
  \to \pi^{0} f_0(980)$ reaction}.
\newblock {\em Phys. Rev. D}, 86:114007, 2012.

\bibitem{Du:2019idk}
Meng-Chuan Du and Qiang Zhao.
\newblock {Internal particle width effects on the triangle singularity
  mechanism in the study of the $\eta(1405)$ and $\eta(1475)$ puzzle}.
\newblock {\em Phys. Rev. D}, 100(3):036005, 2019.

\bibitem{Cheng:2021nal}
Yin Cheng and Qiang Zhao.
\newblock {Hadronic loop effects on the radiative decays of the first radial
  excitations of \ensuremath{\eta} and \ensuremath{\eta}'}.
\newblock {\em Phys. Rev. D}, 105(7):076023, 2022.

\bibitem{Achasov:2015uua}
N.~N. Achasov, A.~A. Kozhevnikov, and G.~N. Shestakov.
\newblock {Isospin breaking decay $\eta(1405) \to f_0(980)\pi^0 \to 3\pi$}.
\newblock {\em Phys. Rev. D}, 92(3):036003, 2015.

\bibitem{Liu:2015taa}
Xiao-Hai Liu, Makoto Oka, and Qiang Zhao.
\newblock {Searching for observable effects induced by anomalous triangle
  singularities}.
\newblock {\em Phys. Lett. B}, 753:297--302, 2016.

\bibitem{Achasov:2021yis}
N.~N. Achasov and G.~N. Shestakov.
\newblock {$\eta$(1295)$\to$ 3 $\pi$ decays}.
\newblock {\em Phys. Rev. D}, 104(11):116026, 2021.

\bibitem{Nakamura:2022rdd}
S.~X. Nakamura, Q.~Huang, J.~J. Wu, H.~P. Peng, Y.~Zhang, and Y.~C. Zhu.
\newblock {Three-Body Unitary Coupled-Channel Analysis on $\eta(1405/1475)$}.
\newblock 12 2022.

\bibitem{Workman:2022ynf}
R.~L. Workman.
\newblock {Review of Particle Physics}.
\newblock {\em PTEP}, 2022:083C01, 2022.

\bibitem{E852:2000rhq}
J.~J. Manak et~al.
\newblock {Partial-wave analysis of the eta pi+ pi- system produced in the
  reaction pi- p ---\ensuremath{>} eta pi+ pi- n at 18-GeV/c}.
\newblock {\em Phys. Rev. D}, 62:012003, 2000.

\bibitem{Stanton:1979ya}
N.~R. Stanton et~al.
\newblock {Evidence for Axial Vector and Pseudoscalar Resonances Near 1.275-GeV
  in eta pi+ pi-}.
\newblock {\em Phys. Rev. Lett.}, 42:346--349, 1979.

\bibitem{BES:1998bgh}
J.~Z. Bai et~al.
\newblock {Partial wave analysis of J / psi --\ensuremath{>} gamma (K+ K-
  pi0)}.
\newblock {\em Phys. Lett. B}, 440:217--224, 1998.

\bibitem{BES:2000adm}
J.~Z. Bai et~al.
\newblock {Partial wave analysis of J / psi ---\ensuremath{>} gamma (K+- K0(S)
  pi-+)}.
\newblock {\em Phys. Lett. B}, 476:25--32, 2000.

\bibitem{Amsler:2004rd}
Claude Amsler et~al.
\newblock {Production and decay of $\eta^\prime$(958) and $\eta$(1440) in
  $\bar{p} p$ annihilation at rest}.
\newblock {\em Eur. Phys. J. C}, 33:23--30, 2004.

\bibitem{BES:1999axp}
J.~Z. Bai et~al.
\newblock {Partial wave analysis of J / psi --\ensuremath{>} gamma (eta pi+
  pi-)}.
\newblock {\em Phys. Lett. B}, 446:356--362, 1999.

\bibitem{E852:2001ote}
G.~S. Adams et~al.
\newblock {Observation of pseudoscalar and axial vector resonances in pi- p
  ---\ensuremath{>} K+ K- pi0 n at 18-GeV}.
\newblock {\em Phys. Lett. B}, 516:264--272, 2001.

\bibitem{OBELIX:1995zjg}
A.~Bertin et~al.
\newblock {E / iota decays to K anti-K pi in anti-p p annihilation at rest}.
\newblock {\em Phys. Lett. B}, 361:187--198, 1995.

\bibitem{CrystalBarrel:1998pap}
A.~Abele et~al.
\newblock {Study of anti-p p --\ensuremath{>} eta pi0 pi0 pi0 at rest}.
\newblock {\em Nucl. Phys. B}, 514:45--59, 1998.

\bibitem{GAMS:1997pxg}
D.~Alde et~al.
\newblock {Partial-wave analysis of the eta pi0 pi0 system produced in pi- p
  charge exchange collisions at 100-GeV/c}.
\newblock {\em Phys. Atom. Nucl.}, 60:386--390, 1997.

\bibitem{CrystalBarrel:1995kfe}
C.~Amsler et~al.
\newblock {E decay to eta pi pi in anti-p p annihilation at rest}.
\newblock {\em Phys. Lett. B}, 358:389--398, 1995.

\bibitem{Anisovich:2000kx}
A.~V. Anisovich, C.~A. Baker, C.~J. Batty, D.~V. Bugg, C.~Hodd, V.~A. Nikonov,
  A.~V. Sarantsev, V.~V. Sarantsev, and B.~S. Zou.
\newblock {A J(PC) = 0-+ enhancement at the f0(980) eta threshold}.
\newblock {\em Phys. Lett. B}, 472:168--174, 2000.

\bibitem{Anisovich:2001jb}
A.~V. Anisovich, D.~V. Bugg, N.~Djaoshvili, C.~Hodd, J.~Kisiel, L.~Montanet,
  A.~V. Sarantsev, and B.~S. Zou.
\newblock {Resonances in anti-p p --\ensuremath{>} eta pi+ pi- pi+ pi- at
  rest}.
\newblock {\em Nucl. Phys. A}, 690:567--594, 2001.

\bibitem{Du:2022nno}
Meng-Chuan Du, Yin Cheng, and Qiang Zhao.
\newblock {Vertex corrections due to the triangle singularity mechanism in the
  light axial vector meson couplings to $K*\bar{K}$+c.c.}
\newblock {\em Phys. Rev. D}, 106(5):054019, 2022.

\bibitem{KLOE:2002kzf}
A.~Aloisio et~al.
\newblock {Study of the decay $\phi \to \eta \pi^0 \gamma$ with the KLOE
  detector}.
\newblock {\em Phys. Lett. B}, 536:209--216, 2002.

\bibitem{KLOE:2002deh}
A~Aloisio et~al.
\newblock {Study of the decay $\phi \to \pi^0 \pi^0 \gamma$ with the KLOE
  detector}.
\newblock {\em Phys. Lett. B}, 537:21--27, 2002.

\bibitem{BESIII:2022chl}
M.~Ablikim et~al.
\newblock {Study of $\eta(1405)/\eta(1475)$ in $J/\psi\to\gamma K^{0}_{S}
  K^{0}_{S}\pi^{0}$ decay}.
\newblock 9 2022.

\end{thebibliography}

\bibliographystyle{unsrt}

\end{document}